\documentclass{aastex631}
\usepackage{CJK}
\usepackage{amsmath}

\begin{document}

\title{Can Neptune's Distant Mean-Motion Resonances Constrain Undiscovered Planets in the Solar System? Lessons from a Case Study of the 9:1}

\author[0000-0002-9989-4782]{Matthew W. Porter}
\affiliation{Department of Physics, University of Michigan\\ Ann Arbor, MI 48109, USA\\}

\author[0000-0001-6942-2736]{David W. Gerdes}
\affiliation{Department of Physics, University of Michigan\\ Ann Arbor, MI 48109, USA\\}
\affiliation{Department of Astronomy, University of Michigan \\
Ann Arbor, MI 48109, USA}

\author[0000-0003-4827-5049]{Kevin J. Napier}
\affiliation{Department of Physics, University of Michigan\\ Ann Arbor, MI 48109, USA\\}

\author[0000-0001-7737-6784]{Hsing~Wen~Lin (\begin{CJK*}{UTF8}{gbsn}
林省文\end{CJK*})}
\affiliation{Department of Physics, University of Michigan\\ Ann Arbor, MI 48109, USA\\}

\author[0000-0002-8167-1767]{Fred C.~Adams}
\affiliation{Department of Physics, University of Michigan\\ Ann Arbor, MI 48109, USA\\}
\affiliation{Department of Astronomy, University of Michigan \\
Ann Arbor, MI 48109, USA}

\begin{abstract}

Recent observational surveys of the outer Solar System provide evidence that Neptune’s distant $n$:1 mean-motion resonances may harbor relatively large reservoirs of trans-Neptunian objects (TNOs). In particular, the discovery of two securely classified 9:1 resonators, 2015 KE$_{172}$ and 2007 TC$_{434}$, by the Outer Solar System Origins Survey is consistent with a population of order $10^4$ such objects in the 9:1 resonance with absolute magnitude $H_r < 8.66$. %\citep{Volk2018}. 
This work investigates whether the long-term stability of such populations in Neptune's $n$:1 resonances can be used to constrain the existence of distant $5-10M_{\oplus}$ planets orbiting at hundreds of AU. The existence of such a planet has been proposed to explain a reported clustering in the orbits of highly eccentric ``extreme" trans-Neptunian objects (eTNOs), although this hypothesis remains controversial. We engage in a focused computational case-study of the 9:1 resonance, generating synthetic populations and integrating them for 1 Gyr in the presence of 81 different test planets with various masses, perihelion distances, eccentricities, and inclinations. While none of the tested planets are incompatible with the existence of 9:1 resonators, our integrations shed light on the character of the interaction between such planets and nearby $n$:1 resonances, and we use this knowledge to construct a simple, heuristic method for determining whether or not a given planet could destabilize a given resonant population. We apply this method to the currently estimated  properties of Planet 9, and find that a large primordial population in the 15:1 resonance (or beyond), if discovered in the future, could potentially constrain the existence of this planet.

\end{abstract}


\section{Introduction and Motivation} \label{sec:intro}
 The gravitational influence of the giant planets displays prominently in the structure of the modern-day Solar System, most notably in small-body populations such as asteroids and trans-Neptunian objects (TNOs). These populations have slowly been sculpted by the planets over their 4.5 billion-year histories, and display prominent structural features such as the large gaps in the semi-major axis distribution of asteroids which correspond to unstable mean-motion resonances with Jupiter \citep{MD1999}. Since exploration of the trans-Neptunian region began \citep{JewittLuu}, structural features in the outer Solar System have inspired may authors to speculate about the existence of potential unseen planets beyond Neptune \citep[see, e.g.,][]{Brown2004, Lykawka&Mukai}. In recent years, anomalous orbital clustering observed in distant TNO populations, first noted by \citet{TS2014}, has been claimed as evidence of an undiscovered 5-10$M_{\oplus}$ planet orbiting the Sun at hundreds of AU \citep{BB2016, P9Rev}. This hypothetical planet is often referred to as Planet 9. The Planet 9 hypothesis remains controversial in recent literature, however, with multiple authors arguing that the seemingly anomalous observations that inspired it are explained instead by selection bias in TNO surveys \citep{Shankman2017, Bernardinelli2020, Napier2021}. With this continuing ambiguity, additional constraints are of interest. 

This work investigates whether the stability of TNO populations occupying distant mean-motion resonances (MMRs) with Neptune could be used to constrain the existence of Planet 9-like objects. We were primarily motivated by the recent discovery of two resonant TNOs, 2015 KE$_{172}$ and 2007 TC$_{434}$, by the Outer Solar System Origins Survey \citep[OSSOS;][]{Volk2018, OSSOSMain, Bannister2018}. \citet{Volk2018} claim to securely classify both objects as occupants of Neptune's 9:1 mean-motion resonance at $a\approx130.1$ AU, and further infer from these discoveries a large population of 9:1 resonators of order $\sim 10^4$ objects with $H_r < 8.66$. With eccentricities of $\sim0.67$, these orbits extend to aphelion distances of $Q\sim220$ AU, potentially into the realm of Planet 9's dynamical influence. Since Planet 9's orbit and mass are not precisely constrained according to current literature (\citealt{BB2021, BB2022}; see also \citealt{P9Rev} for a detailed assessment), we seek to answer the following questions: (1) Do Planet 9-like objects destabilize, or otherwise affect the dynamics of distant $n$:1 resonances with Neptune? If so, how strongly, and in what manner? (2) Given a known set of resonant objects, how would one approach placing quantitative constraints on the allowed properties of additional planets? (3) Could the small sample of known 9:1 resonators be used to place meaningful constraints on Planet 9's orbit? We employ computational methods to investigate these questions through a focused case study of the 9:1 resonance, generating and integrating synthetic resonant populations in the presence of various Planet 9-like objects. \citet{Clement2021} performed a similar suite of integrations on several distant $n$:1 resonances, using a single set of Planet 9 parameters. Their results showed little action for resonances nearer than 12:1, but the specific version of Planet 9 used in their simulations is both less massive and more distant than has been suggested in more recent literature \citep[e.g.][]{BB2021, BB2022}. 

The remainder of Section~\ref{sec:intro} gives a brief overview of the dynamics of the 9:1 resonance, while Section~\ref{sec:methods} describes our methods for generating and integrating synthetic resonators. Section~\ref{sec:noP9} characterizes the long-term stability of the 9:1 resonance assuming that no undiscovered planets exist, while Section~\ref{sec:effects} illustrates the effects of Planet 9-like perturbers on the resonance. Lastly, Section~\ref{sec:arbitrary} extrapolates our result to the case of a general $n$:1 resonance, and outlines a crude, first-glance method for constraining planet properties from the assumption of a stable population at arbitrary values of $n$.

\subsection{Background: Neptune's $n$:1 resonances}
For a given integer $n$, the $n$:1 resonance with Neptune can occur when the semi-major axis $a$ is such that the orbital period is $n$ times that of Neptune \citep{MD1999}. Following from Kepler's third law, the requisite semi-major axis can be expressed succinctly:
\begin{equation}
    a_{n:1} = a_{N}n^{2/3} \approx n^{2/3} \cdot 30.07 \; \textup{AU}
\end{equation}
where $ a_{N}$ is Neptune's semi-major axis. It is important to note that proximity to this semi-major axis value is a \textit{necessary}, but not a \textit{sufficient} condition for an object to truly be classified as resonant. Mean-motion resonance is a dynamical phenomenon: over many successive orbits, resonant objects will undergo a periodic libration in both semi-major axis as well as another parameter, called the \textit{resonant angle}, which we denote in this work as $\phi$ \citep{Saillenfest2020}. The definition of $\phi$ varies for different MMRs, but for $n$:1 resonances it can generally be expressed as
\begin{equation}
    \phi = n\lambda - \lambda_{N} - (n-1)\varpi \,,
\end{equation}
where $\lambda$ is the resonator's mean longitude, $\lambda_{N}$ is Neptune's mean longitude, and $\varpi$ is the resonator's longitude of perihelion. We note that other forms of the resonant angle exist for $n$:1 resonances, but for most purposes this one is most important \citep{Saillenfest2020}, and will be the focus of this investigation. 

\begin{figure}[ht]
    \centering
    \hspace*{-1cm}\includegraphics[scale = 0.5]{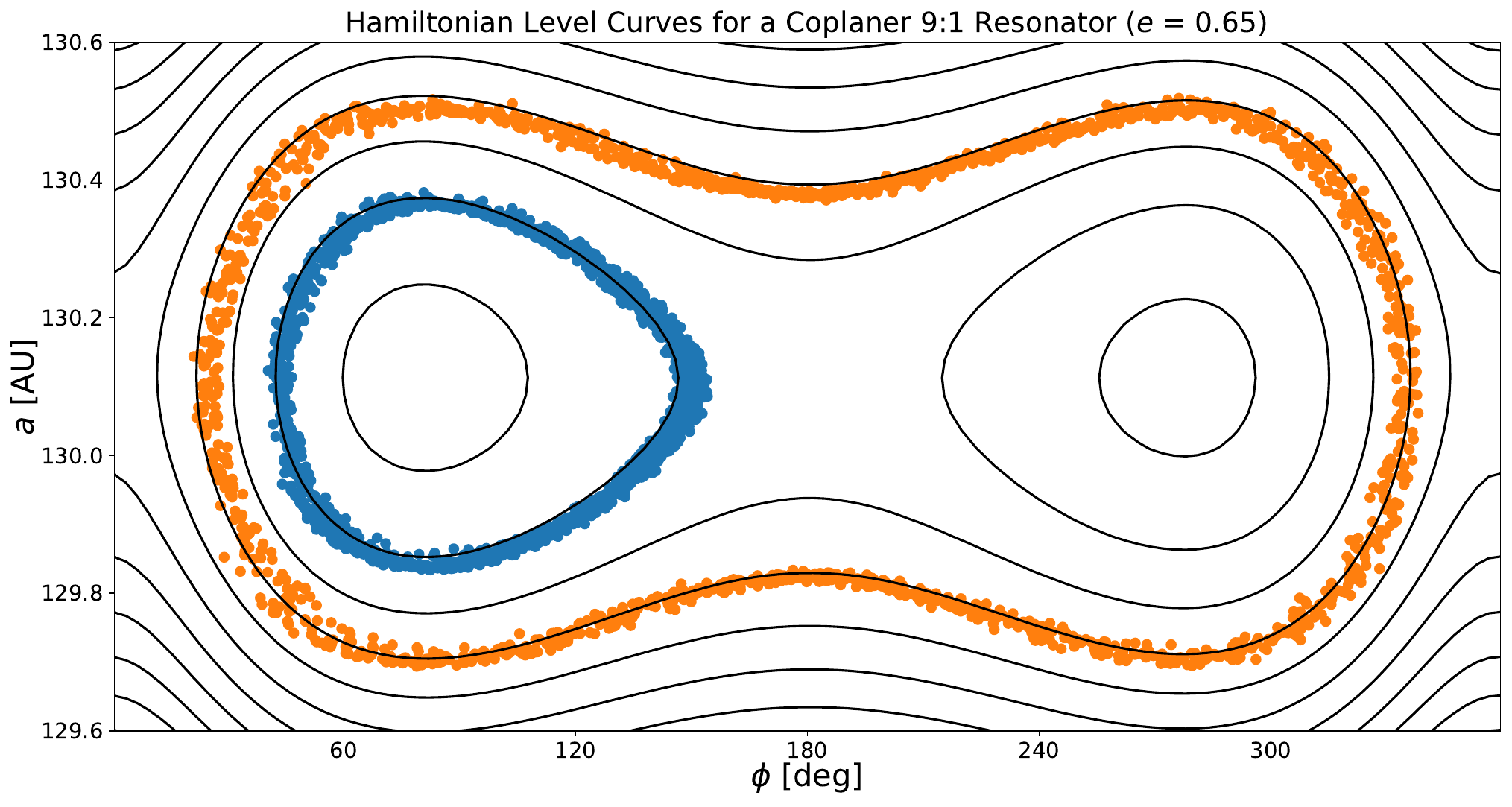}
    \caption{Level curves of the resonant Hamiltonian (eq. 3) for a zero-inclination 9:1 resonator ($e = 0.65$). The evolution of two synthetic resonators over a short 3 Myr $n$-body integration is also plotted. One of these objects (blue dots) was placed in the leading asymmetric island, while the other (orange dots) was placed in the symmetric island. Both objects initially had $e = 0.65$ and $i = 0$. The curves are plotted in $a$ rather than $\Sigma$, as $\Sigma$ is approximately linear in $a$ over the small range of resonant behavior.}
    \label{fig:hamiltonian}
\end{figure}

The resonant libration follows the level curves of the Hamiltonian (given in \citealt{Saillenfest2020}) in the ($\phi$, $a$) - plane:
\begin{equation}
    \mathcal{H}_{res} = -\frac{\mu^2}{2(n\Sigma)^2} - \dot{\lambda}_N \Sigma - \sum_{i = \textup{J, S, U}} \frac{\mu_i}{4 \pi^2} \int_0^{2\pi} \int_0^{2\pi} \frac{1}{\lVert \textbf{r} - \textbf{r}_i \rVert} \textup{d} \lambda_i \textup{d} \lambda - \frac{\mu_N}{2\pi} \int_0^{2\pi} \left( \frac{1}{\lVert \textbf{r} - \textbf{r}_N \rVert} - \frac{\textbf{r} \cdot \textbf{r}_N}{\lVert \textbf{r}_N \rVert^3}\right) \textup{d}{\gamma}\,.
\end{equation}
Here $\dot{\lambda}$, $\dot{\lambda}_N$ denote the mean motion of the resonator and Neptune respectively, $\mu$ is the standard gravitational parameter of the Sun, and $\Sigma$ is a new coordinate defined according to 
\begin{equation}
    \Sigma \equiv \frac{\sqrt{\mu a}}{n}\,,
\end{equation}
where the integer $n=9$ for most cases of interest here. The first summation, the \emph{secular} term, is a double average of the gravitational potential over the orbital motion of the resonator and each of the giant planets, excluding Neptune. The following term is the \emph{resonant} term, and is a single average over the correlated orbital motion of Neptune and the resonator. Note that this Hamiltonian, and thus the resulting shape of the level curves, is a function of the resonator's orbital elements. A plot of the level curves for a coplanar 9:1 resonator with $e$ = 0.65 ($q$ = 45.5 AU) is shown in Figure~\ref{fig:hamiltonian}.
These curves display three distinct libration modes: two \textbf{asymmetric} islands centered approximately at $\phi = 90 ^{\circ}$ and $270^{\circ}$, and a \textbf{symmetric} island encircling them centered at $\phi = 180^{\circ}$. The $90^{\circ}$ asymmetric island is typically referred to as the \emph{leading} island, while the one at $270^\circ$ is called \emph{trailing}. This structure is a shared feature across all of the external $n$:1 resonances \citep[see, e.g,][]{Gladman2012}, but the precise shape of the Hamiltonian level curves will vary depending on a particular resonator's eccentricity and inclination.

\section{Goals and Methods} \label{sec:methods}
We seek to study the long-term stability of 9:1 resonators over Gyr timescales in the presence of Planet 9-like objects. As only two 9:1 resonators have been discovered to date, we proceed by generating and integrating synthetic resonant populations. For each of the candidate planets we test, we generate an initial population of 1000 resonators. The procedure we used to generate these objects was chosen due to the lack of data on the real resonant population. The obvious solution to this constraint is to simply populate the resonant parameter space uniformly, but this approach introduces the further problem of outputting disproportionately many highly-inclined orbits, which are relatively unlikely in the real Solar System. Thus, when generating our objects, we use a more observationally motivated inclination distribution: $\sin{(i)}$ multiplied by a gaussian of width 30 degrees. This functional form was first used by \citet{Brown2001} to model the inclination distribution of scattered disk objects. More recently, \citet{Cromp} used this same form, with gaussian widths of 20-25 degrees, to model the inclinations of objects in distant $n$:1 resonances with Neptune. We adopt this same form, with a slightly broader width, to bias our study toward more reasonable orbits while still exploring a broad range of inclinations. The other orbital elements are generated uniformly over the ranges given in Table~\ref{tab:generation}.

\begin{figure}[ht]
    \centering
    \hspace*{-1cm}\includegraphics[scale = 0.5]{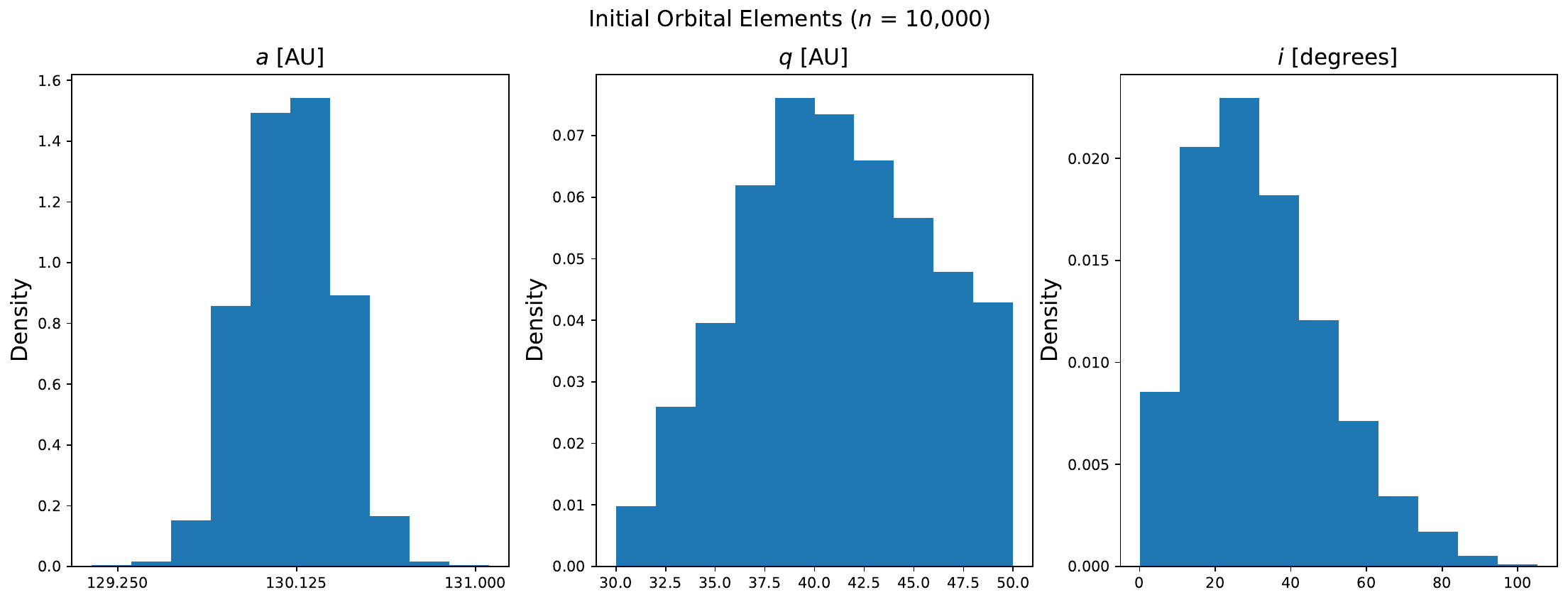}
    \caption{Initial distribution of orbital elements resulting from the generation process. Note that while these distributions result from the case of no additional perturber, the perturber had no effect on these distributions over 10 Myr. Also note that these distributions are not the same as those given in Table \ref{tab:generation}, since all objects which did not librate in the 9:1 for at least 10 Myr were discarded.}
    \label{fig:element_dist}
\end{figure}

\begin{table}[ht]
    \centering
    \begin{tabular}{|c|c|}
    \hline
    \textbf{Orbital Element} & \textbf{Generating Distribution}
    \\
    \hline
    \textbf{Semi-major axis} ($a$) & Uniform from 129.1 to 131.1 AU
    \\
    \hline
    \textbf{Perihelion distance} ($q$) & Uniform from 30 to 50 AU
    \\
    \hline
    \textbf{Inclination} ($i$) & $\sin{(i)}$ times gaussian ($\sigma = 30^{\circ}$)
    \\
    \hline
    \textbf{Longitude of ascending node} ($\Omega$) & Uniform from 0 to 360$^{\circ}$
    \\
    \hline
    \textbf{Longitude of perihelion} ($\varpi$) & Uniform from 0 to 360$^{\circ}$
    \\
    \hline
    \textbf{Mean Anomaly} & Uniform from 0 to 360$^\circ$
    \\
    \hline
    \end{tabular}
    \caption{Generating distributions used in the first step of the resonator generation process. Generated orbits are integrated for 10 million years and discarded if unstable.}
    \label{tab:generation}
\end{table}
Our full population generation process is as follows: (1) Sample an orbit from the given element distributions. (2) Integrate the orbit for 10 million years in the presence of the four giant planets and the additional perturber. (3) Discard the object if the resonant behavior is unstable over 10 million years. (4) Repeat until 1000 short-term stable resonators are obtained. Figure~\ref{fig:element_dist} shows the initial distributions of semi-major axis, perihelion distance, and inclination that result from this process. Note that these initial distributions are non-uniform because of the refinement in step 3 that discards highly unstable objects, even though the orbits were sampled from uniform distributions.

We test 81 different versions of the distant planet, varying its mass, perihelion distance, eccentricity, and inclination over a 3$\times$3$\times$3$\times$3 grid in the parameter space. The particular values we test, given in Table \ref{tab:P9_params}, are inspired by recent literature on the Planet 9 hypothesis. \citet{BB2021} prefer a planet with a mass of $6.2^{+2.2}_{-1.3}M_{\oplus}$, a semi-major axis of $380^{+140}_{-80}$ AU, a perihelion of $300^{+85}_{-60}$ AU, and an inclination of $16 \pm 5^{\circ}$, though we note these authors have since claimed to rule out a portion of this parameter space using observational data \citep{BB2022}. For mass, perihelion, and inclination, we use their central and 1-$\sigma$ values, choosing eccentricity to be one of 0.2, 0.3, or 0.4. All planets we test are assigned $\varpi = 248^{\circ}$ and $\Omega = 98^{\circ}$, the central values reported by \citet{BB2021}. A separate set of 1000 resonators is generated for all 81 combinations of mass, perihelion, eccentricity, and inclination. For the remainder of this work, we refer to the distant planet candidates as \emph{perturbers}.

\begin{table}[ht]
    \centering
    \begin{tabular}{|c|c|c|c|}
    \hline
    \textbf{Parameter} & \textbf{Low Value} & \textbf{Intermediate Value} & \textbf{High Value}
    \\
    \hline
    \textbf{Mass ($M_{\oplus}$)} & 4.9 & 6.2 & 8.4
    \\
    \hline
    \textbf{Perihelion (AU)} & 240 & 300 & 385
    \\
    \hline
    \textbf{Eccentricity} & 0.2 & 0.3 & 0.4
    \\
    \hline
    \textbf{Inclination} & 11$^\circ$ & 16$^\circ$ & 21$^\circ$
    \\
    \hline
    \end{tabular}
    \caption{Tested values of the perturber parameters. All combinations of $m$, $q$, $e$, and $i$ were tested, for a total of 81 perturbers. The given $q$-$e$ combinations produce perturbers with semi-major axis of 300.0 AU, 342.9 AU, 375.0 AU, 400.0 AU, 428.6 AU, 481.3 AU, 500.0 AU, 550.0 AU, and 641.7 AU}
    \label{tab:P9_params}
\end{table}

After the initial populations are obtained, we integrate them for 1 Gyr in the presence of the additional perturber, treating the resonators as test particles. For these integrations, we use \texttt{rebound}'s WHFast integrator \citep{Rein2012} through the \texttt{spacerocks} Python package \citep{Napier_spacerock} with a timestep of 100 days. Included in each integration are the four giant planets, the additional perturber, and the 1000 resonators generated as described. Data was written out every 100,000 years due to disk space constraints, which is insufficient to fully resolve the resonant libration. In order to properly classify the test particles at the end of the 1 Gyr integrations, we performed short 2 Myr extension integrations on all test particles with final semi-major axes within 1 AU of the resonance. These extension integrations included the four giant planets as well as the additional perturber, and the state vector was written out every 2,000 years so that the test particles could be checked for resonant libration at high time resolution. As a control, we also generated, integrated, and classified 10,000 objects using the same process, but with no additional perturber. In order to save computational resources, we considered removing objects from the 1 Gyr integration after they leave the 9:1 resonance. However, we were interested in checking how these objects diffuse into the wider Solar System, so we instead only removed objects that crossed $a < 0$ or $a > 1000$ AU, which we considered to be ejected from the Solar System.

It is insightful to further discuss the perihelion distance interval used the initial step of the generation process. While the lower cutoff at $q = 30$ AU is physically motivated by the presence of Neptune's orbit at this distance, the upper cutoff at $q = 50$ AU is more arbitrary. Indeed, \citet{DistantDynamics} showed that resonant behavior can occur at surprisingly high perihelion distances, and simple cuts in $q$ are insufficient to accurately represent the boundaries of Neptune's resonances. To investigate the 9:1 at high perihelion distances, we sampled 5000 orbits from the distributions from Table~\ref{tab:generation}, but with the upper perihelion cutoff extended to $q = 100$ AU. These orbits were integrated for 10 Myr and checked resonant libration. We find that the range of semi-major axes where libration occurs declines roughly exponentially with increasing perihelion, diminishing rapidly above $q = 50$ AU, although we find small number of librating particles out to $q = 100$ AU.

It is clear from these integrations that the range of resonant behavior can extend well above our $q = 50$ AU generation cutoff, albeit over a rapidly diminishing width in $a$. We found that most of the 10 Myr-stable objects (70\%) fall below this perihelion threshold. Furthermore, the two 9:1 objects reported by \citet{Volk2018} both had $q < 50$ AU. The authors find these discoveries consistent with a model population of \~10,000 objects with $q$ uniformly between 39 and 52 AU, noting that OSSOS would not have been sensitive to objects with higher perihelia. For the purposes of this analysis, it follows that a $q = 50$ AU generation cutoff explores most of the resonant parameter space while efficiently focusing on objects that would be detectable to current surveys.
\section{Characterizing the Unperturbed Resonance} \label{sec:noP9}
\begin{figure}[ht]
    \centering
    \hspace*{-1cm}\includegraphics[scale = 0.55]{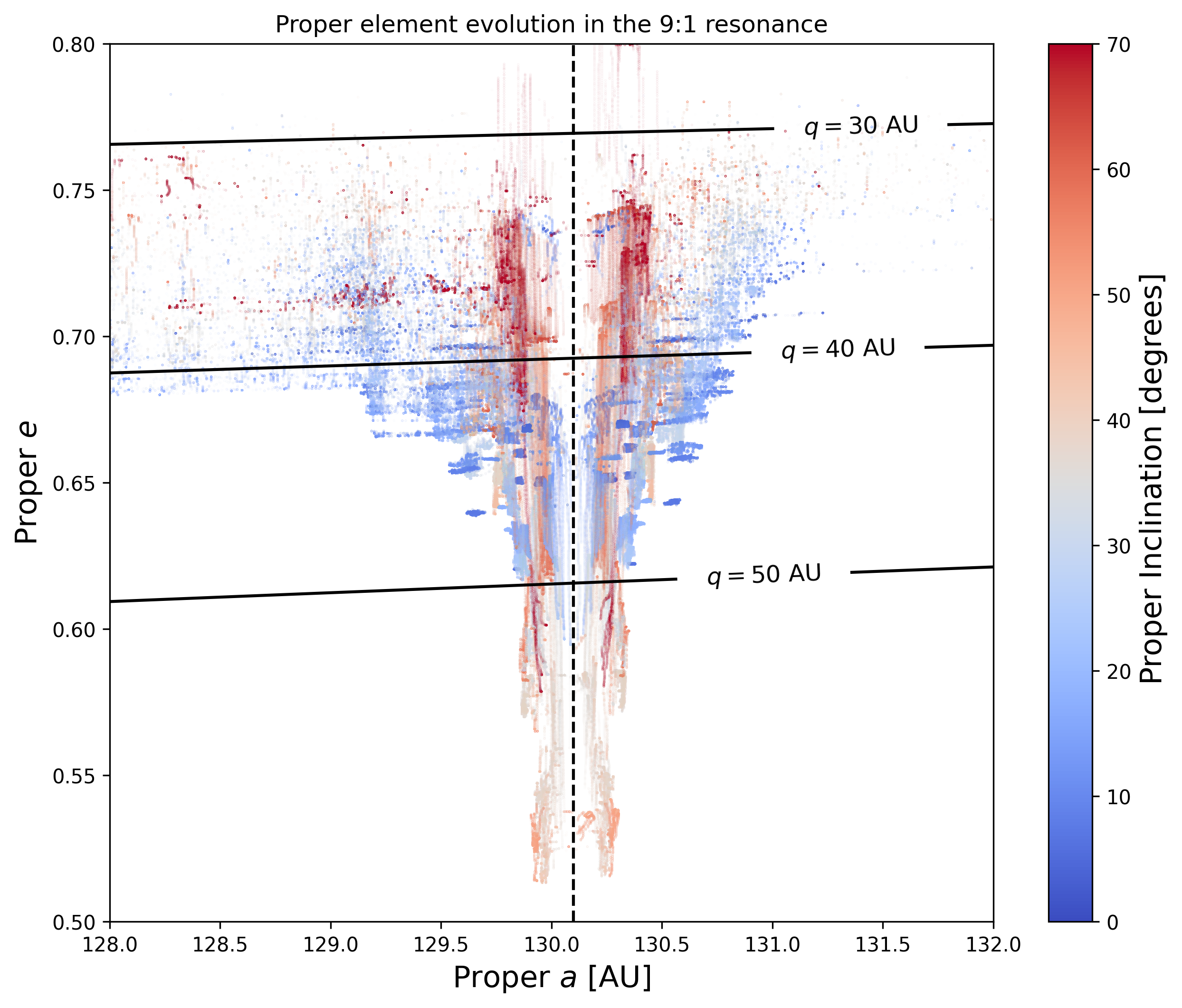}
    \caption{Evolution of proper orbital elements for 100 randomly selected objects from the control simulations, which included only the four known giant planets and the generated resonators. While proper $a$ is defined as the maximum of $a$ over a sliding 10 Myr window, in this figure we follow \citet{Morb1997} and plot each object's sliding maximum \emph{and} minimum $a$, in order to preserve visual symmetry about the center of the 9:1 resonance. The dashed vertical line indicates the exact center of the resonance at 130.1 AU, while the solid black lines are curves of constant perihelion distance.}
    \label{fig:ResStructure}
\end{figure}
Before discussing the effects of the perturber, it is useful to examine the results of the control simulations in detail. Past studies of the stability of Neptune's resonances \citep[e.g., ][]{Morb1997, Tisc2009} approach such analysis by defining \emph{proper orbital elements}, in order to tease out long-term dynamical behavior from the quasi-periodic oscillations induced by the resonance. We follow this approach by defining the proper $a$, $e$, and $i$ to be the maximum value of the orbital element in question over a sliding 10 Myr window. Similar to the analysis of \citet{Morb1997}, The proper semi-major axes of our test particles exhibit four distinct dynamical outcomes when no additional planet is present. Firstly, the particle's proper $a$ may exhibit no measurable diffusion, indicating a highly stable orbit; secondly, it may diffuse slowly, but not escape the 9:1 resonance over the 1 Gyr integration time; thirdly, it may exhibit slow diffusion followed by an escape from the resonance after several hundred Myr; or fourthly, it may diffuse rapidly, exiting the resonance almost immediately.

Further analysis shows that the 9:1 resonance has a discernible dynamical structure when viewed in the plane of proper $a$ and $e$, as illustrated in Figure \ref{fig:ResStructure}. Most orbits with $q \lesssim 40$ AU are catastrophically unstable and quickly diffuse out of the 9:1 resonance. This is not unexpected; \citet{Saillenfest2020} showed that resonant TNOs with perihelia near Neptune are strongly influenced by planetary scattering, and \citet{GLISSE} observed a similar long-term depletion of low-$q$ objects in their study of Neptune's 3:1 resonance. Highly stable orbits are found at low inclinations and high perihelion distances ($q \gtrsim 40$ AU), visible as dark blue patches in Figure \ref{fig:ResStructure}. Some of the objects at moderate to high inclination undergo Kozai oscillations, periodically cycling to low eccentricities and gaining inclination as they do so. Kozai inside MMRs is not a new result: \citet{Morb1997} found the Kozai resonance to play an important role in the structure of Neptune's 3:2 exterior resonance, while \citet{Gomes2005} explored this effect as a mechanism for populating the high-perihelion scattered disk. In the 9:1, objects experiencing Kozai oscillations may exit the resonance if their eccentricity cycling brings them into the highly diffusive region below $q \approx 40$ AU. 
\begin{figure}[ht]
    \centering
    \hspace*{-1cm}\includegraphics[scale = 0.4]{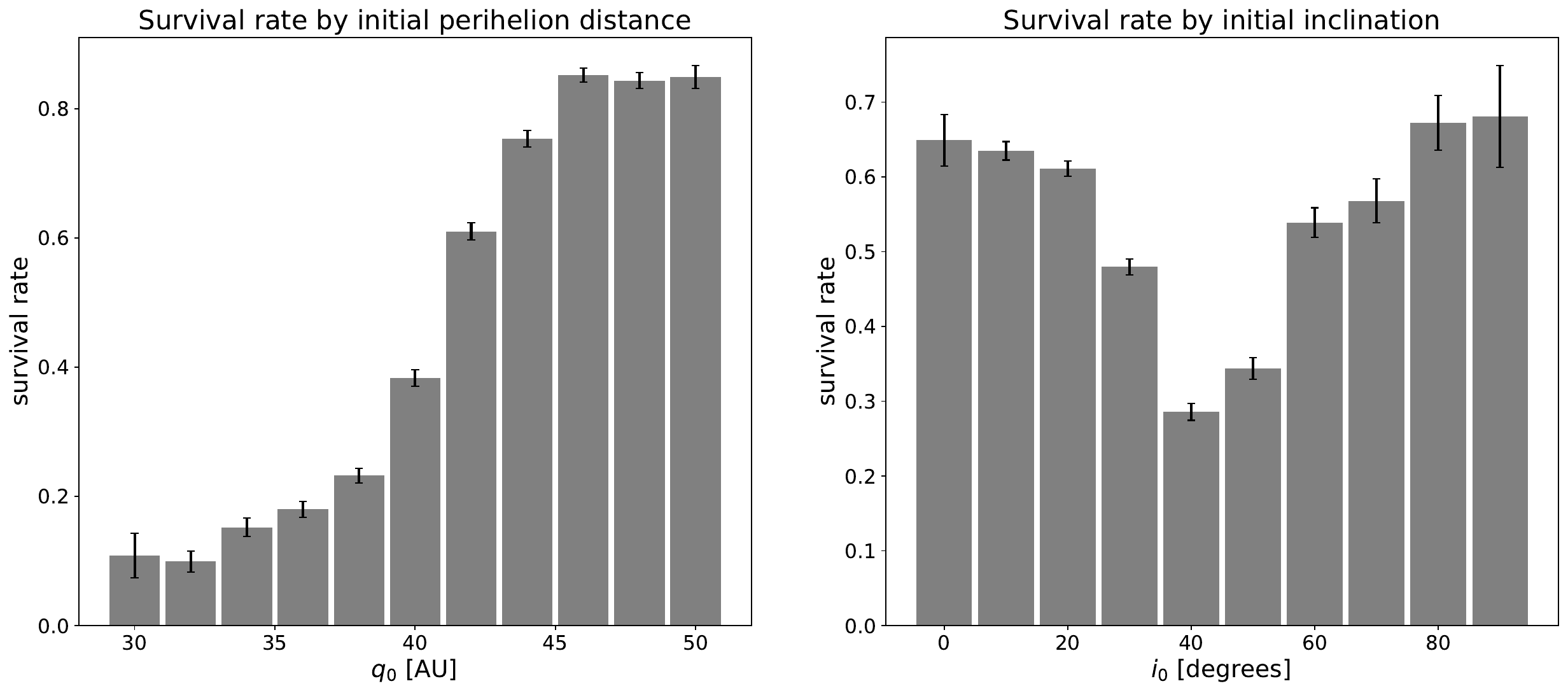}
    \caption{Survival rate as a function initial perihelion distance (left) and inclination (right) in the control simulations, which contained only the four known giant planets with no distant perturber. Inclination bins are centered on multiples of 10$^\circ$, and span $5^\circ$ in either direction of the center. Perihelion distance bins are centered on multiples of 2 AU, and span 1 AU in both directions of the center. 1$\sigma$ Error bars are derived by treating the number of surviving objects in each bin as a Poisson variable. %The poor statistics at high inclination result from the low number of high-$i$ objects produced by our generation process.
    }
    \label{fig:survival}
\end{figure}

At the end of the 1 Gyr integration, we classified the objects as \emph{9:1 resonant}, \emph{non-9:1 resonant}, or \emph{ejected} based on both their final semi-major axes and their behavior during the extension integration. An object was considered \emph{9:1 resonant} if the resonant angle remained bounded on the interval $0 < \phi < 355^{\circ}$ for the entire 2 Myr extension integration, indicating stable libration. Similarly, \emph{non-9:1 resonant} refers to any non-ejected particle that either did not meet the libration criteria during the extension integration, or was not included in the extension integration at all, due to its final semi-major axis being more than 1 AU from the center of the resonance. Finally, \emph{ejected} objects were those that were removed during the main 1 Gyr integrations, due to leaving the semi-major axis interval $0<a<1000$ AU. A large fraction of the 10,000 objetcs in the control simulation were stable, with 41.4\% being classified as 9:1 resonant after 1 Gyr. 50.3\% of the objects ended the simulation as non-9:1 resonant, while only 8.3\% had been ejected from the Solar System.

\begin{figure}[ht]
    \centering
    \hspace*{-1cm}\includegraphics[scale = 0.6]{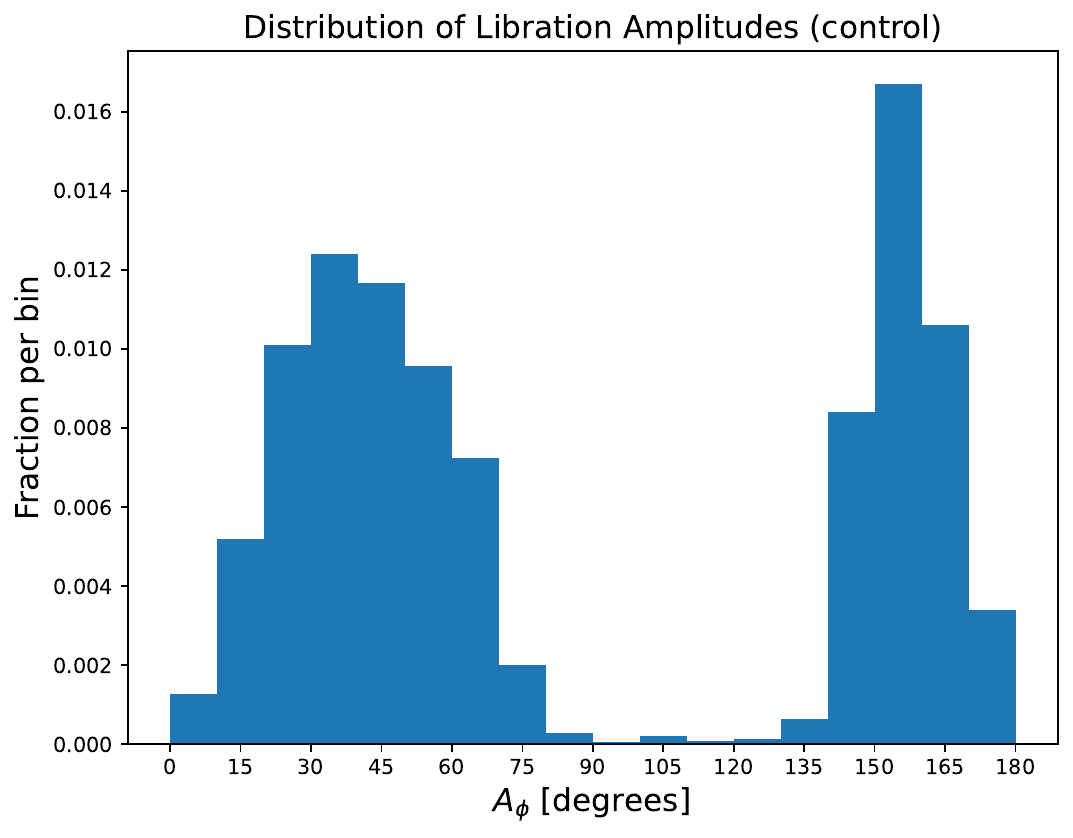}
    \caption{Normalized distribution of libration amplitude in the 2 Myr extension of the control integration. Of the 10,000 total objects in the control simulation, 4,138 exhibited libration in the extension integration.}
    \label{fig:A_dist}
\end{figure}

After 1 Gyr, the surviving resonators are predominantly high-perihelion objects, with over 80\% having $q > 40$ AU. Long-term resonant stability is observed across a broad range of inclinations; in particular, we note the presence of two highly stable regions near $i = 0$ and $i = 90^{\circ}$, while inclinations near $45^{\circ}$ are less stable. Figure \ref{fig:survival} shows the rate at which objects survive the 1 Gyr integration as resonant, as a function of initial perihelion distance and inclination. It also also interesting to examine the final distribution of libration amplitudes among the resonant objects, shown in Figure \ref{fig:A_dist}. We computed this quantity directly from the output of the 2 Myr extension integration, defining the amplitude as $A_{\phi} = \frac{1}{2}(\textup{max}(\phi) - \textup{min}(\phi))$.

The bimodal structure of Figure \ref{fig:A_dist} is expected from the structure of the Hamiltonian. The population with $A_{\phi} < 90^{\circ}$ is composed of objects librating in one of the two asymmetric islands, while the group having $A_{\phi} > 90^{\circ}$ represents the symmetric librators. A modest majority of librating objects (59.8\%) were found to be in one of the two asymmetric islands, with neither the leading nor trailing island hosting significantly more objects than the other. By treating the numbers of asymmetric and symmetric librators as Poisson variables, we calculate the ratio of asymmetric librators to symmetric librators to be 1.49 +/- 0.05. This crude prediction must be taken with a grain of salt, however. The precise shape of the Hamiltonian, and thus the relative sizes of the symmetric and asymmetric islands, are sensitive to both eccentricity and inclination. The value of this ratio in the real Solar System could thus vary substantially depending on the underlying demographics of the  9:1 population, even in the absence of a distant perturber. Still, the observed prevelance of asymmetric librators is not surprising; \citet{Yu2018} numerically studied the sticking times of scattering objects in Neptune's MMRs, and found that asymmetric librators tend to be longer lived than their symmetric counterparts in $n$:1 resonances.

\section{Effects of the Perturber} \label{sec:effects}
\begin{figure}[ht]
    \centering
    \hspace*{-1cm}\includegraphics[scale = 0.5]{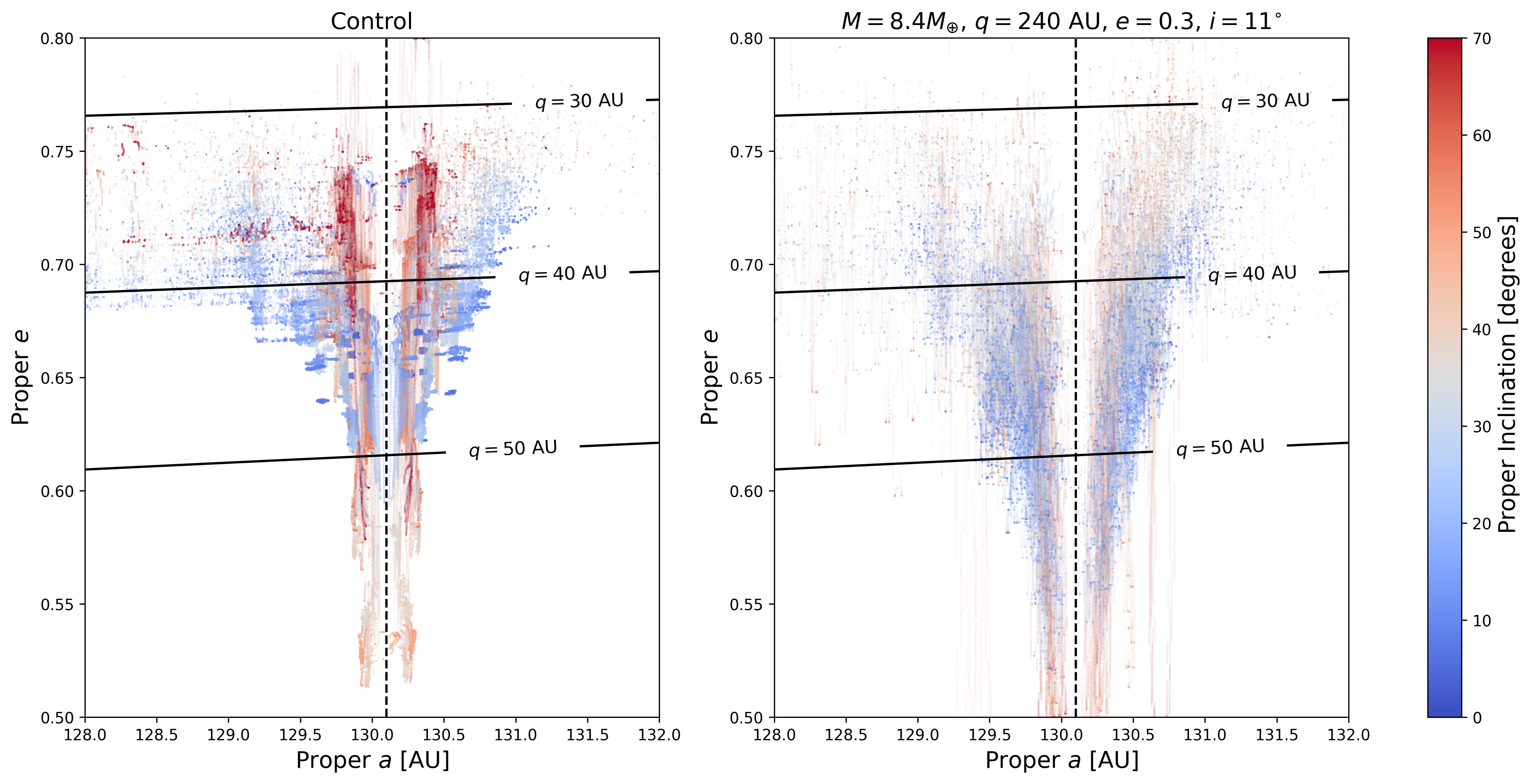}
    \caption{Evolution of proper orbital elements for 100 randomly selected objects from the integration involving the perturber with $M = 8.4M_{\oplus}$, $q = 240$ AU, $e = 0.3$, and $i = 11^{\circ}$. As with Figure \ref{fig:ResStructure}, both the maximum and minimum of $a$ over the sliding 10 Myr window are plotted for each object to preserve visual symmetry. The dashed verticle line indicates the center of the 9:1 resonance, while the solid black lines are curves of constant perihelion distance.}
    \label{fig:ResComp}
\end{figure}
The distant perturbers we tested displayed a broad range of dynamical effects. The perturber's influence was similar in character across the tested masses and orbital elements, but was generally quite weak in all but the most aggressive cases. As such, the discussion and illustrations in this section will largely be limited to the cases in which the perturber had the highest mass ($m = 8.4M_{\oplus}$) and lowest perihelion distance ($q = 240$ AU), in order to best display the effects.  In addition, we found that the strength of these effects varied only mildly over the range of tested perturber inclinations and eccentricities, with $m$ and $q$ being more important. As a result, for each of the mass-perihelion distance combinations in Table~\ref{tab:P9_params}, we grouped the objects from the 9 corresponding eccentricity-inclination combinations together for analysis in order to obtain larger samples. Since we generated 1,000 resonators for each perturber, each of these larger samples contains 9,000 particles.

Figure \ref{fig:ResComp} shows how the perturber with $M = 8.4M_{\oplus}$, $q = 240$ AU, $e = 0.3$, and $i = 11^{\circ}$ alters the resonant structure in the plane of proper $a$ and $e$. In contrast to the control simulations, we did not observe the same class of highly stable, low-$i$ objects above $q = 40$ AU; all objects in this simulation instead underwent large fluctuations in proper $a$, $e$, and $i$, with most diffusing out of the resonance altogether. The effects of the perturbers on the stability and structure of the resonance are analyzed in detail below: Section~\ref{sec:adiff} discusses how the perturbers change the overall survival rate of resonators over the full 1 Gyr, while Section~\ref{sec:qdiff} identifies and discusses the mechanism by which the perturber ejects objects from the resonance. Lastly, Section~\ref{sec:lowamp} illustrates the perturbers' effects on how objects occupy the three different libration modes shown in Figure~\ref{fig:hamiltonian}.
\subsection{Diffusion of Semi-major Axis} \label{sec:adiff}
\begin{figure}[ht]
    \centering
    \hspace*{-0.75cm}\includegraphics[scale = 0.45]{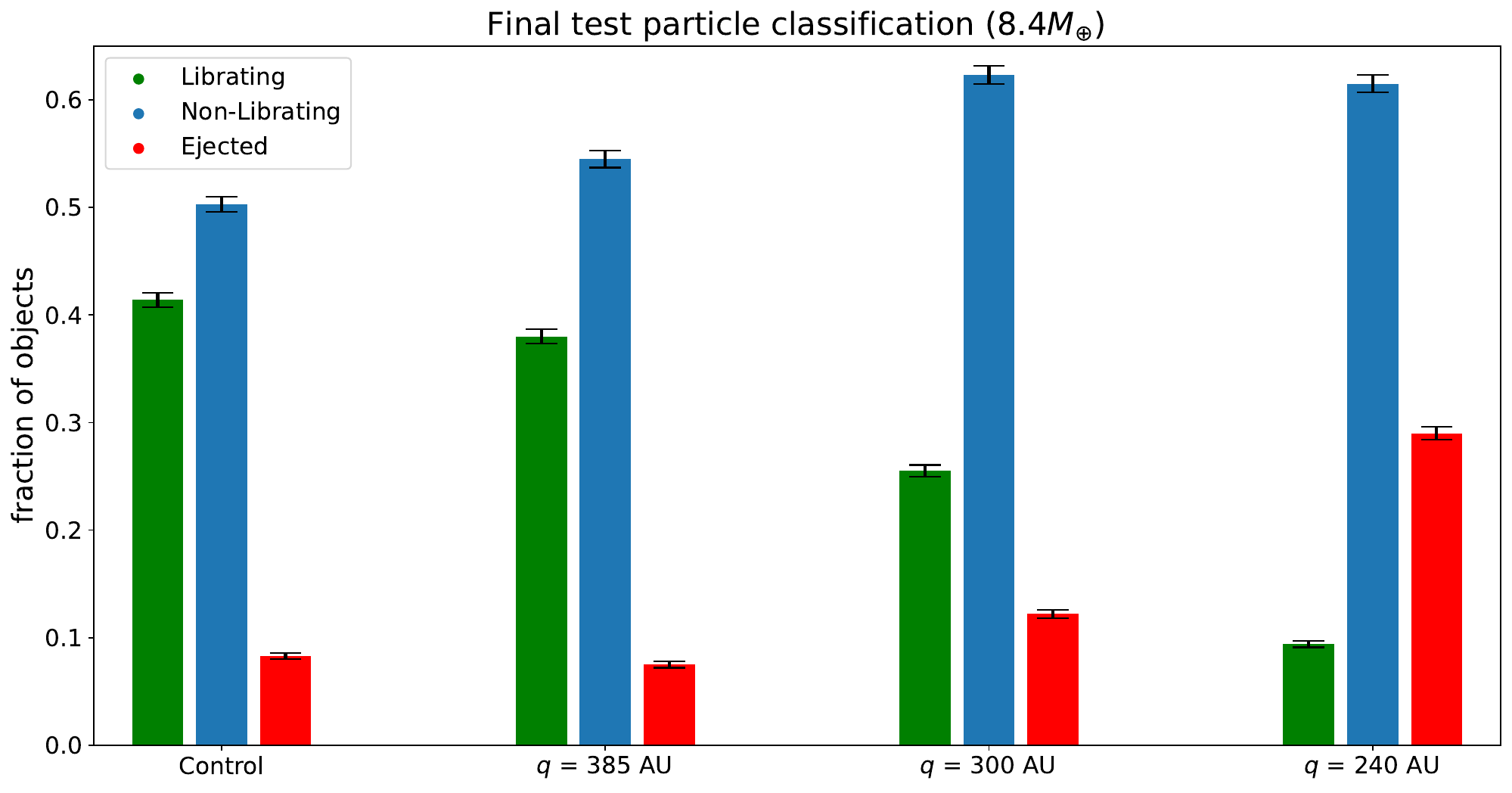}
    \caption{Final semi-major axis classification of objects for perturbers with different perihelion distances. All perturbers had a mass of 8.4$M_{\oplus}$. For each each perihelion distance, we group the 9 perturber eccentricities and inclinations together into a single sample in order to obtain better statistics. Error bars ($1\sigma$) are derived by treating the number of objects in each classification category as a Poisson variable.}
    \label{fig:classification}
\end{figure}
\begin{figure}[ht]
    \centering
    \hspace*{-0.75cm}\includegraphics[scale = 0.55]{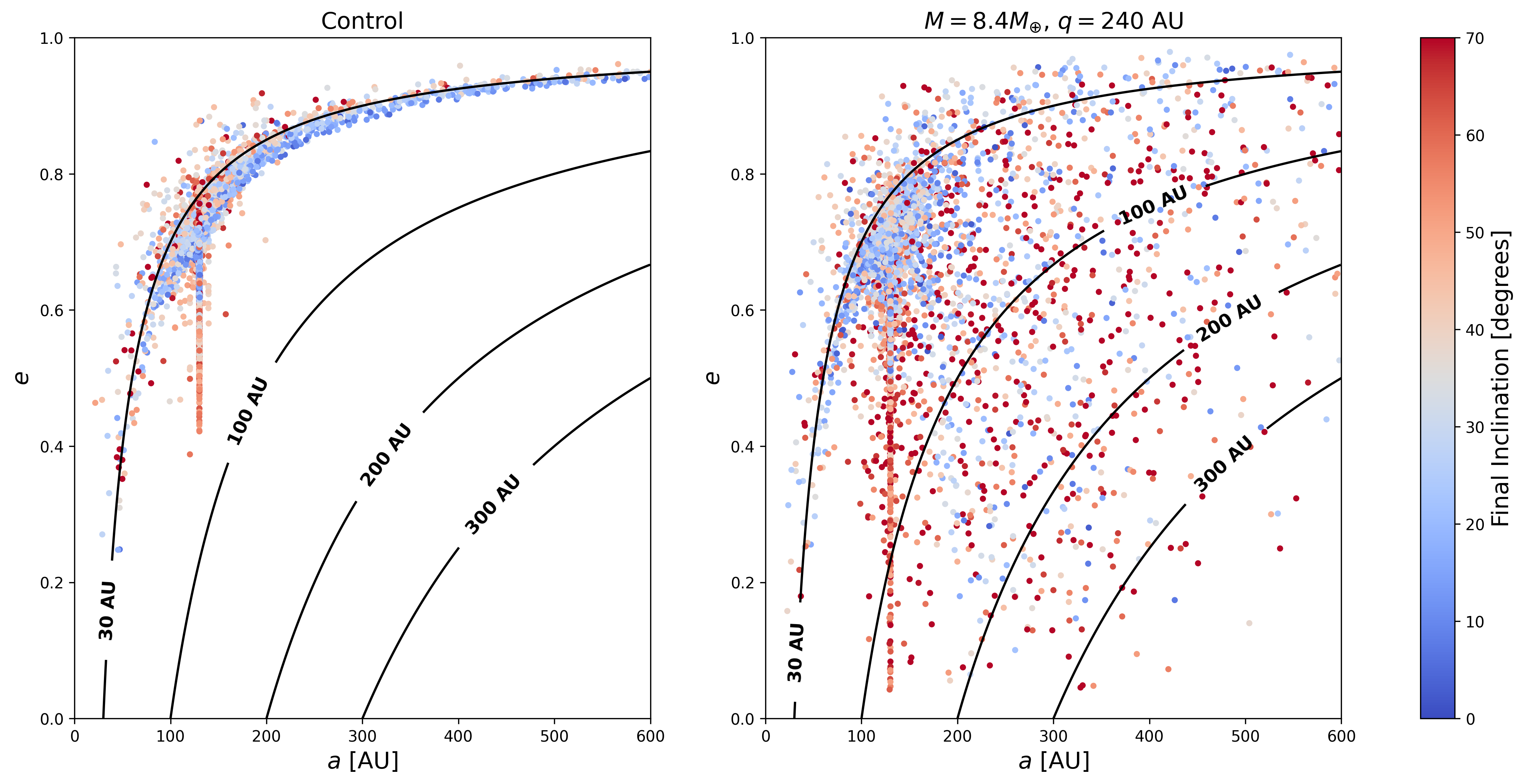}
    \caption{Final semi-major axis and eccentricity of objects after 1 Gyr, for all 10,000 objects in the control simulation (left), and all 9,000 objects in simulations involving an $8.4M_{\oplus}$, $q = 240$ AU perturber (right). Colors indicate final inclination, and the black contours are curves of constant perihelion. In the control case, most objects that exit the 9:1 remain at low inclinations perihelion distances, while objects affected by a perturber are raised to high $q$ and $i$}
    \label{fig:a_vs_e}
\end{figure}
The simplest effect of the additional perturber is the overall reduction in the number of objects that remain in the 9:1 resonance for the entire integration. When the classification system discussed in Section~\ref{sec:noP9} is applied to the simulations with a perturber, a modest but significant decline in 9:1 resonant objects is observed in most cases, with the $q = 240$ AU, $m = 8.4M_{\oplus}$ perturbers causing particularly large declines. We find that this decrease in the resonant population is accompanied primarily by an increase in the number of \textit{ejected} objects, while the incidence of \textit{non-9:1 resonant} objects varies modestly in comparison. Figure~\ref{fig:classification} compares the final classification from several simulations to the control population characterized in Section~\ref{sec:noP9}.

While \textit{ejected} objects constitute a largely self-explanatory category, it is interesting to ask how the \textit{non-9:1 resonant} objects occupy the wider Solar System across these various cases. In particular, it is important to ask whether these objects are preferentially captured into long-term stable libration in other mean-motion resonances, either with Neptune or with the perturber, after they are removed from the 9:1. If this is the case, it would suggest that a distant perturber could act to drastically \textit{reorder} the structure of Neptune's distant resonant populations rather than to merely destabilize them, implying a suite of much more complicated effects that this analysis, which focuses on a single resonance, could not model. A simple visual inspection of Figure~\ref{fig:a_vs_e}, however, is sufficient to show that long-term stable resonance recaptures are infrequent at best. As shown in the right-hand panel, most objects that the perturber ejects from the 9:1 have their perihelia lifted to values high enough to protect them from planetary scattering ($q \gtrapprox 40$ AU), rendering resonance capture with Neptune largely impossible. This is consistent with the findings of \citet{Clement2021}, who observed that Planet 9 exerts a perihelion-raising effect on scattering objects in the vicinity of Neptune's distant resonances.

Qualitatively, the final population of non-9:1 resonant objects is significantly different under the influence of an $8.4M_{\oplus}$, $q = 240$ perturber when compared to the control case. In addition to the large perihelion raising effect, we observed that the perturber greatly increases the population of high-$i$ and retrograde objects. In the combined set of simulations involving an $8.4M_{\oplus}$, $q = 240$ AU perturber, 6.6\% of non-9:1 resonant objects ended with $i > 75^{\circ}$, compared to only 2.1\% in the control sample. For retrograde orbits, the occurence rates were 2.5\% ($8.4M_{\oplus}$, $q = 240$ AU perturber), versus 0.26\% (control). The maximum final inclination of any non-ejected particle was $173.7^{\circ}$ ($8.4M_{\oplus}$, $q = 240$ AU perturber), versus $106.1^{\circ}$ (control). This is consistent with the findings of past literature on the distant giant planet hypothesis: both \citet{Shankman2017b} and \citet{Lawler2017} find that an eccentric, inclined super-Earth exerts a significant perihelion-raising effect on large-$a$ TNOs, and also frequently raises them to large or even retrograde inclinations. \citet{Becker2018} similarly find that a distant giant planet can help explain the existence of high-inclinaion eTNOs such as 2015 BP$_{519}$.
\subsection{Perihelion Cycling: the Resonance Kickout Mechanism} \label{sec:qdiff}
\begin{figure}[ht]
    \centering
    \hspace*{-0.75cm}\includegraphics[scale = 0.5]{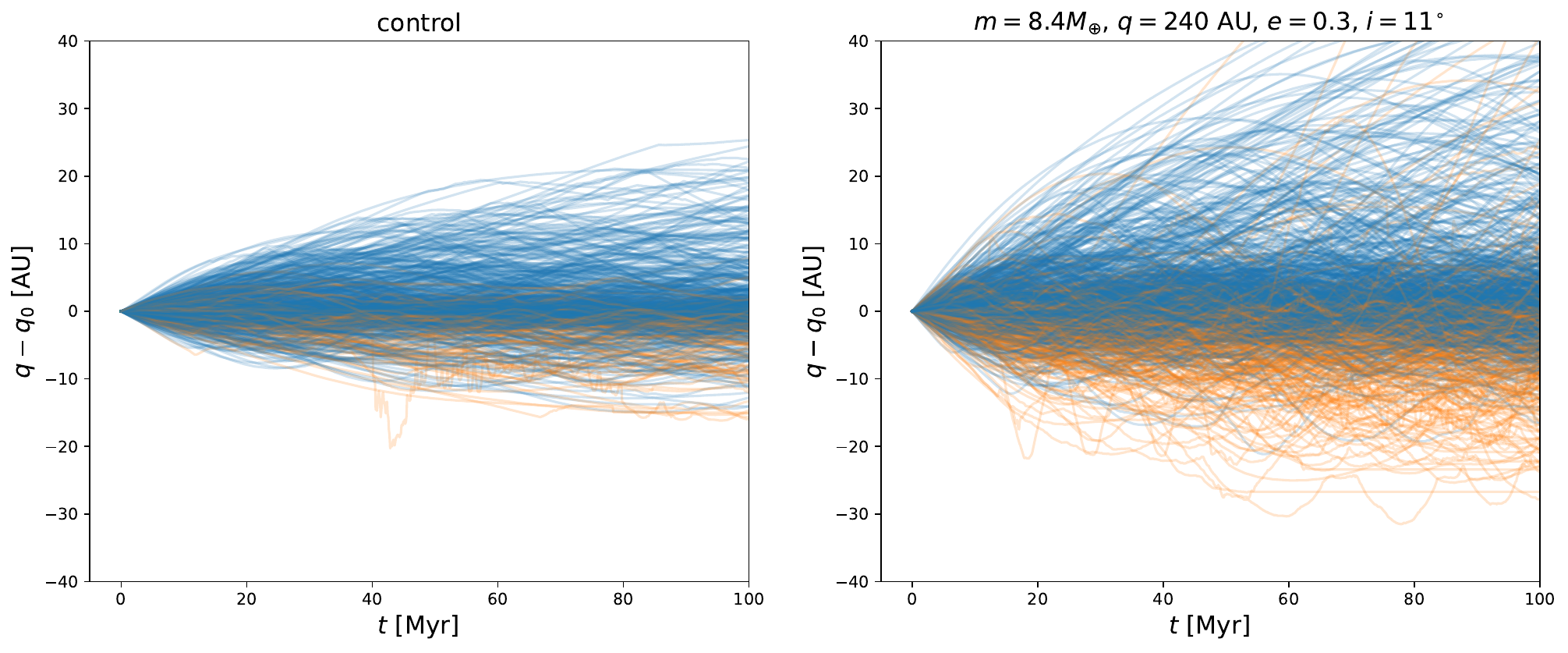}
    \caption{Deviation of perihelion distance from initial value for 1,000 objects in one of the control simulations (left), and in the simulation with the $m$ = 8.4$M_{\oplus}$, $q$ = 240 AU, $e$ = 0.3, $i$ = 11$^\circ$ perturber (right). Objects with semi-major axes within 1 AU of the center of 9:1 resonance at $t = 100$ Myr are colored in blue, while those with semi-major axes more than 1 AU distant from the resonance center at this time are colored in orange. Perihelion distances diffuse much faster in the latter case due to the perturber.}
    \label{fig:q_diff}
\end{figure}
\begin{figure}[ht]
    \centering
    \hspace*{-0.75cm}\includegraphics[scale = 0.5]{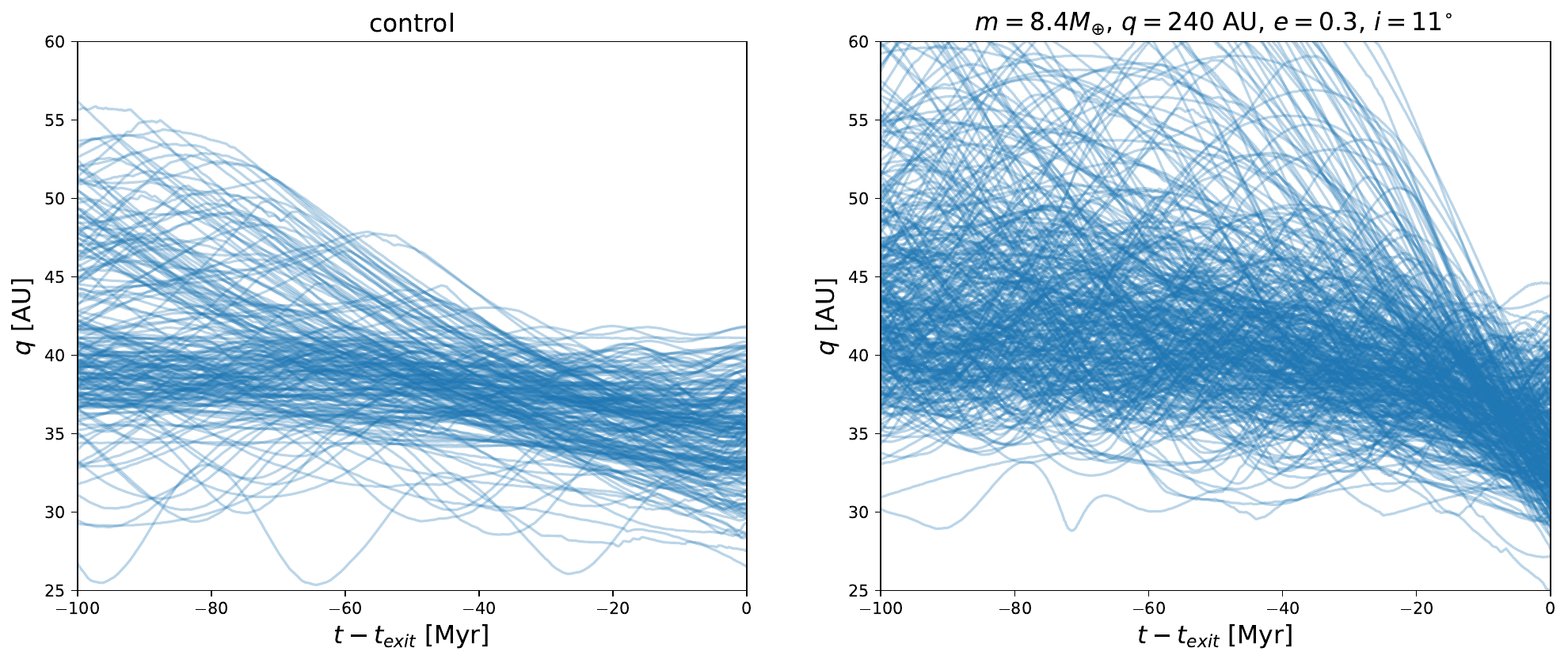}
    \caption{Perihelion evolution during the 100 Myr prior to exiting the interval 129.1 AU < $a$ < 131.1 AU, for all interval-leaving objects in one of the control simulations (left), and in the simulation with the $M = 8.4M_{\oplus}$, $q$ = 240 AU, $e$ = 0.3, $i$ = 11$^\circ$ (right). Clear negative trends feature in both plots, illustrating that the ejected objects are those driven into the low-$q$ scattering region. Far more objects leave resonance in the latter case, primarily because the perturber causes more objects to enter this unstable region.}
    \label{fig:q_kickout}
\end{figure}
The $q = 240$ AU, $m = 8.4M_{\oplus}$ perturbers induced dramatic changes in the objects' perihelion distances over the 1 Gyr integrations. This effect is illustrated by Figure~\ref{fig:q_diff}, which plots the deviation from the initial $q$ value over the first 100 Myr for all 1000 objects in two of the simulations. In the control case, only highly inclined objects experienced large oscillations in $q$, while low-inclination objects had stable perihelia. In contrast, the perturbing planet provides an accessible reservoir of angular momentum across all inclinations; all of the test particles in the integrations involving a perturber experienced substantial perihelion cycling. Discussion of this effect is especially pertinent in light of the features shown in Figure~\ref{fig:survival}, as the dropoff in survival rate at low perihelion distances implies that objects driven to such $q$ values by the perturber's influence are in danger of being ejected from the resonance. 

In order to test whether this is the main mechanism causing resonance kickout, we examined how the perihelion distances of the objects change shortly before exiting the interval 129.1 AU < $a$ < 131.1 AU. For all objects that left this interval, we examined the change in $q$ during the 100 Myr leading up to that object's interval exit time $t_{\textup{exit}}$. Figure~\ref{fig:q_kickout} shows the resulting perihelion trajectories for all objects that exit this interval in two of the simulations, clearly demonstrating that this effect is indeed the primary mechanism driving resonance kickout. In the perturber simulation shown in Figure~\ref{fig:q_kickout}, for instance, objects that left the interval had a mean perihelion value of $\bar{q}_{\textup{exit}} = $ 33.97 AU at the time of exit, while the mean perihelion distance for the remaining objects at the end of the simulation was $\bar{q} = $ 51.76 AU. For comparison, in the control simulation on the left of Figure~\ref{fig:q_kickout}, these values were $\bar{q}_{\textup{exit}} = $ 35.46 AU (interval-leaving) and $\bar{q} = $ 45.69 AU (remaining). Survival is clearly highly dependent on $q$ in both cases, but the perturber causes far more objects to enter the unstable low-$q$ region than otherwise would. It is also not surprising that the mean perihelion distance of objects which remain in this interval is significantly higher when an aggressive perturber is present, as Figure~\ref{fig:q_diff} shows that the perturber's influence causes significant perihelion diffusion in both directions. The objects driven to low $q$ values are eliminated from the resonance, leading to a net raising of $\bar{q}$ for the remaining objects. The perturber can even cause some objects to reach such high values of $q$ that Neptune's influence becomes extremely weak, in some cases leading to the cessation of resonant behavior despite remaining at the commensurate semi-major axis $a \approx 130.1$ AU. A small population of such objects arose in nearly all simulations, and increased in size with the aggressiveness of the perturber.
\subsection{Erosion of the Asymmetric Libration Islands} \label{sec:lowamp}
\begin{figure}[ht]
    \centering
    \hspace*{-1cm}\includegraphics[scale = 0.5]{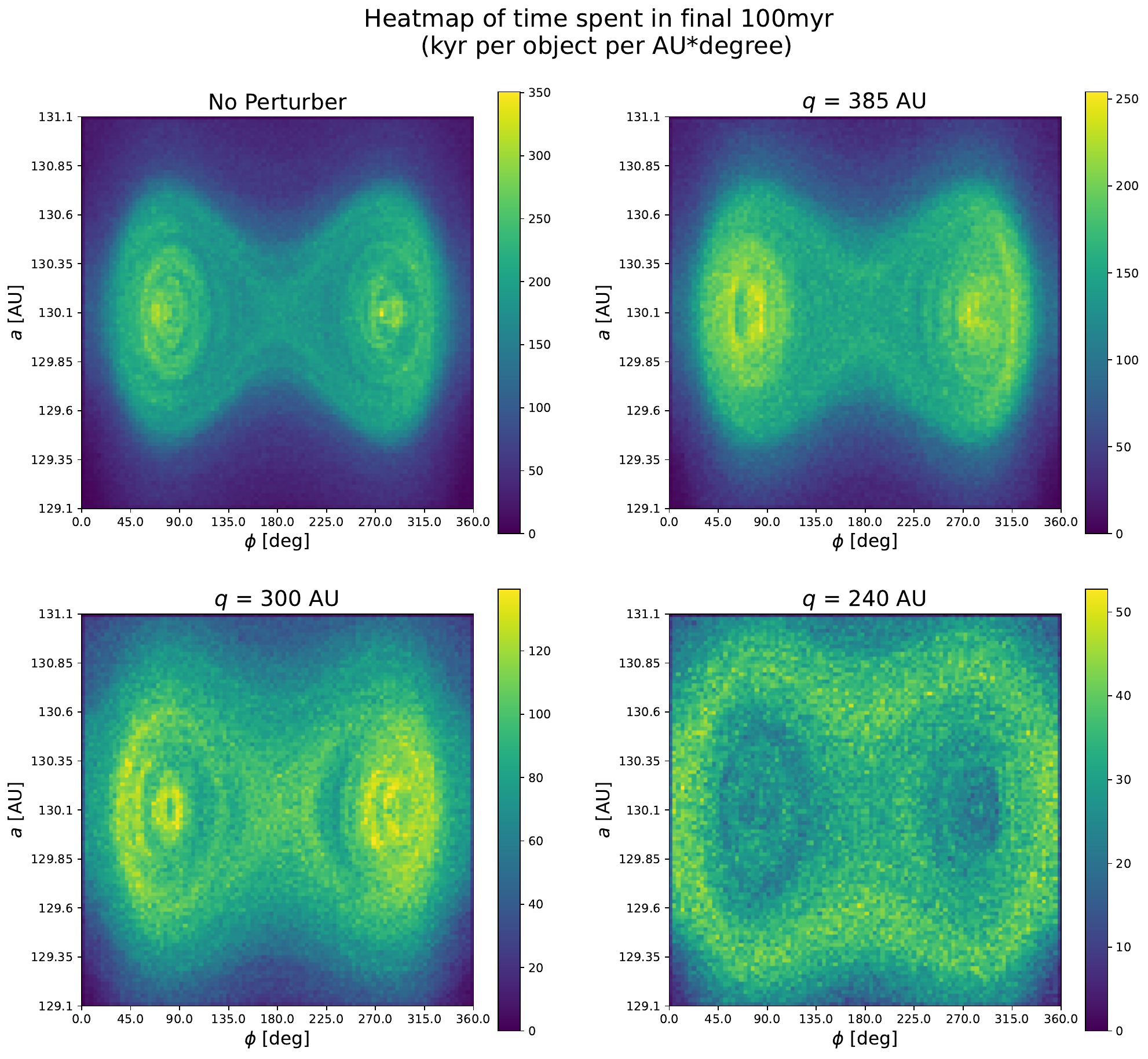}
    \caption{Heatmaps showing how resonators occupy the ($a$, $\phi$) parameter space during the the final 100 Myr of the simulations involving the 8.4 $M_{\oplus}$ planets, in the case of: (top left) No Planet, (top right) $q$ = 385 AU, (bottom left) $q$ = 300 AU, (bottom right) $q$ = 240 AU. For each case, we group the 9 different eccentricity-inclination combinations together as one sample to reduce noise, as these parameters have less impact than $q$ and mass. Note we only included objects with $q < 50$ AU, to remove contamination from dynamically detached objects.}
    \label{fig:heatmaps}
\end{figure}

In addition to examining the reduction of the resonant population, it is also insightful to examine how the additional perturber alters the dynamics of the objects that remain in resonance at the end of the simulation. In particular, it can be asked if the perturber alters how objects are distributed among the three libration islands. To visualize this, we impose a $100\times100$ grid on the parameter space from $129.6$ AU $< a < 130.6$ AU and $0 < \phi_{9:1} < 360^\circ$, counting the number of data points in each grid cell across all objects during the last 100 Myr of the simulation. The results can then be presented as heatmaps, as shown in Figure~\ref{fig:heatmaps} for several representative simulations. Each of the three perturber $q$ values in Figure \ref{fig:heatmaps} represents a sample of 9,000 total particles, obtained by grouping together the simulations representing of each of the nine $e-i$ combinations for the corresponding $8.4M_{\oplus}$ perturber.
In the case of no additional perturber, the heatmap clearly resembles the shape of the Hamiltonian level curves in Figure~\ref{fig:hamiltonian}, with two strong peaks at the inner islands. These peaks are progressively eroded as the perihelion distance of the perturber approaches the aphelion distance of the resonators ($Q \approx 210-220$ AU). We discuss a likely dynamical explanation for this phenomenon in Section~\ref{sec:arbitrary}. As in Section \ref{sec:noP9}, we can directly compute the ratio of asymmetric to symmetric librators for each of the cases shown in Figure \ref{fig:heatmaps} using the 2 Myr extension integrations. As before, we treat the observed number of asymmetric and symmetric librators as Poisson variables to derive error bars. In the control case, this ratio is 1.49 +/- 0.05; for the $q = 385$ AU group (upper right panel), it is 1.34 +/- 0.05; for the $q = 300$ AU group (lower left panel), it is 1.09 +/- 0.05; for the $q = 240$ AU group (lower right panel), it is 0.74 +/- 0.05.

\section{Extrapolating to arbitrary planets and resonances}
\label{sec:arbitrary}
\begin{figure}[ht]
    \centering
    \hspace*{-0.75cm}\includegraphics[scale = 0.385]{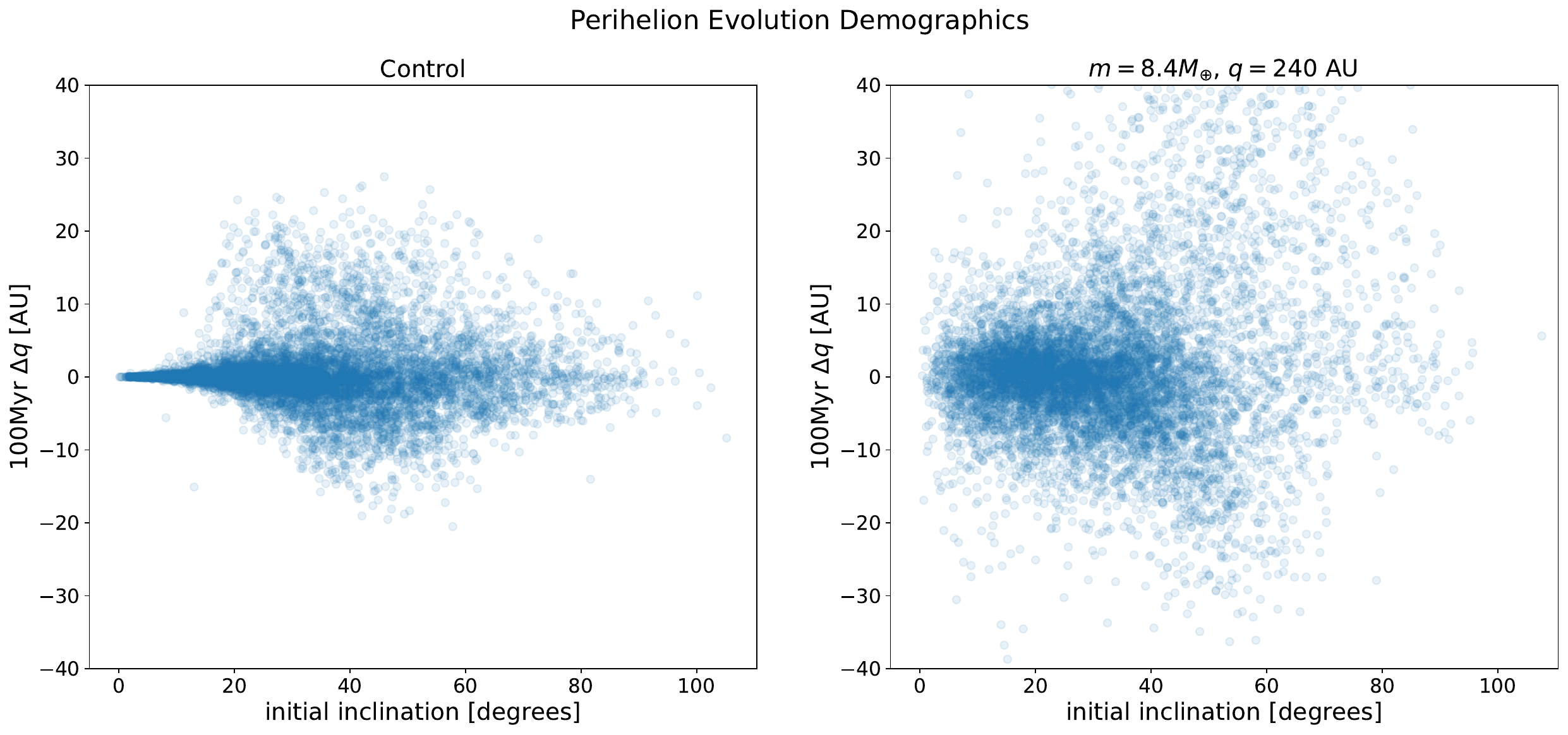}
    \caption{Initial inclination of integrated objects versus signed change in perihelion distance after 100 Myr. (Left): Control simulations. (Right): $m = 8.4M_{\oplus}$ $q = 240$ AU perturber, aggregated over all 9 eccentricity-inclination combinations.}
    \label{fig:q_demographics}
\end{figure}

The previous section demonstrates that additional planets in the outer Solar System can exert significant dynamical influence on $n$:1 Neptune-resonant populations over Gyr timescales, even if those planets do not appreciably alter the structure of the resonant Hamiltonian shown in Figure~\ref{fig:hamiltonian}. These simulations indicate that the mechanism acting to remove objects from resonance occurs as the additional planet drives their perihelion distances to low, unstable values. It follows that a large primordial population of objects in a given $n$:1 resonance with Neptune may be used to place constraints on the existence of nearby massive planets. In order to formulate such constraints, however, a deeper understanding of the mechanism at play is required. Consider the specific case of a low-inclination 9:1 resonator, with a perihelion distance of 45 AU ($e \approx 0.65$), librating deep in one of the asymmetric inner islands in Figure~\ref{fig:hamiltonian}. As demonstrated in Section~\ref{sec:noP9}, these characteristics are emblematic of resonators stable for Gyr-timescales. To understand what planet parameters a large population of such objects would constrain, we ask the following question: \emph{Which planets could drive this object's perihelion distance to an unstable value}? In response to this question, it is useful to think in terms of angular momentum rather than orbital elements. Given a (roughly) constant semi-major axis value, such as in the case of a mean-motion resonance, an object's eccentricity (and thus its perihelion distance) represents a proxy for the magnitude of its angular momentum. Viewed through this lens, planets serve as large reservoirs of angular momentum that resonators can draw from in order to change their perihelion distances. It is clear from this perspective why high-$q$, low-$i$ resonators are so stable in the canonical Solar System: A zero-inclination resonator is prevented from exchanging angular momentum with the giant planets because of the symmetries of their (roughly) circular, co-planar orbits. This fact is embodied by the time-averaged Lagrange planetary equation for $\textup{d}e/\textup{d}t$,
\begin{equation}
    \frac{\textup{d}e}{\textup{d}t} = -\frac{\sqrt{1-e^2}}{\dot{\lambda}a^2e} \frac{\partial \langle \mathcal{R} \rangle}{\partial \varpi}\,,
    \label{eqn:dedt}
\end{equation}
where $\langle \mathcal{R} \rangle$ is the time-averaged disturbing function \citep{MD1999}. In the case of a resonator that is coplanar with the giant planets, $\partial \langle \mathcal{R} \rangle/\partial\varpi$ is \emph{always} equal to 0. Thus, with near-constant perihelion distances that lie well beyond the scattering region, such objects can easily remain trapped in resonance for billions of years. A central feature of the Planet 9 hypothesis, however, is that Planet 9's orbit must be both moderately eccentric and inclined in order to explain the anomalous TNO properties that inspired it \citep{P9Rev}. This feature is important here, as the resulting orbital asymmetries make the angular momentum of Planet 9 much more accessible to nearby resonators than that of the other planets. Figure~\ref{fig:q_demographics} illustrates this fact by comparing how the initial inclinations of our integrated resonators are correlated to the changes in perihelion distance after 100 Myr in two of the simulations. In the left panel, which represents the canonical Solar System with no additional planets, a large population of objects with long-term stable perihelion distances is seen at low inclinations. Moderately inclined objects show much more dramatic changes in $q$, as they are able to exchange angular momentum with Neptune through interactions such as Kozai oscillations. The right panel, on the other hand, shows the combined set of objects from all simulations involving an additional $8.4M_{\oplus}$ perturber with $q = 240$ AU, aggregated over the nine eccentricity-inclination combinations. The additional perturber largely destroys the demographics observed in the left panel. All of the objects now have highly variable perihelion distances due to the new reservoir of easily accessible angular momentum, and are thus at risk of being driven into the scattering region.

With the knowledge that the eccentricity and inclination of the to-be-constrained planet are important, we now return to the central constraining question posed earlier. Existing constraints on additional planets, such as those from infrared surveys or planetary ephemerides, have historically been considered (at least as a starting point) within the plane of planet mass $m$ and orbital distance $r$ \citep[see, e.g.,][]{P9Rev, Silsbee2018, Luhman2014}. However, the importance of the aforementioned orbital asymmetries makes constraints from resonance stability considerably more complex; a distant planet on a circular orbit would be more difficult to exchange angular momentum with than one on an eccentric, inclined orbit. To move forward, the approach described below can be used to assess the risk that such planets pose to nearby resonators, on a planet-by-planet basis.

\begin{figure}[ht]
    \centering
    \hspace*{-0.75cm}\includegraphics[scale = 0.385]{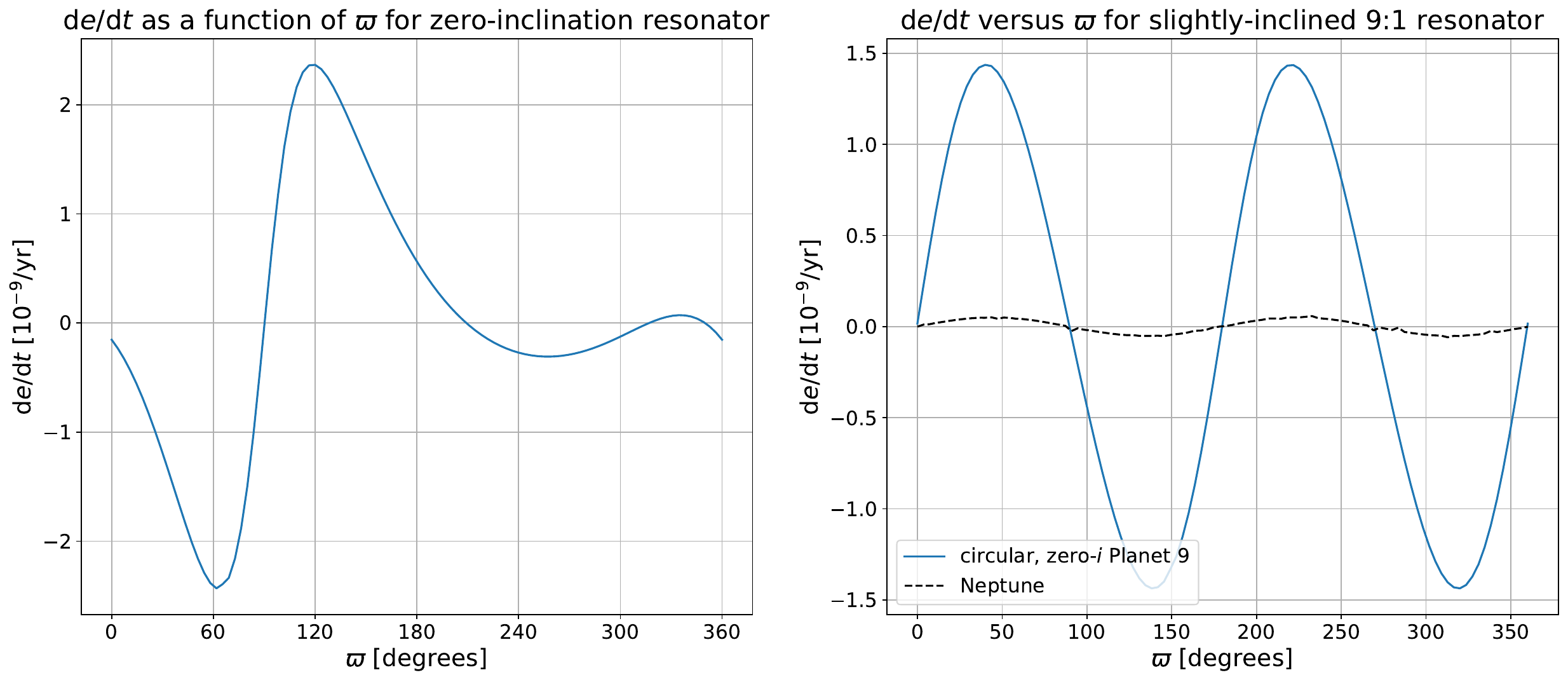}
    \caption{(Left): Computed eccentricity time derivative for a zero-inclination 9:1 resonator with $q = 45$ AU, in the presence of a perturbing planet ($m = 8.4M_{\oplus}$, $a = 400$ AU, $q = 240$ AU, $i = 16^{\circ}$, $\omega = 150^{\circ}$, $\Omega = 97^{\circ}$), as a function of $\varpi$. (Right): Computed contribution to the secular eccentricity time derivative for a 9:1 resonator with $i=5^{\circ}$ from Neptune (black), and a zero-$e$, zero-$i$ planet with $a=240$ AU. The computation was done using equation (5). Note that the contribution to d$e$/d$t$ from this ``Planet 9" is much higher than that from Neptune, \emph{even though} both orbits are circular and coplanar. This is due to the much larger size of this planet's orbit}
    \label{fig:dedt_curve}
\end{figure}

\begin{figure}[ht]
    \centering
    \hspace*{-0.75cm}\includegraphics[scale = 0.38]{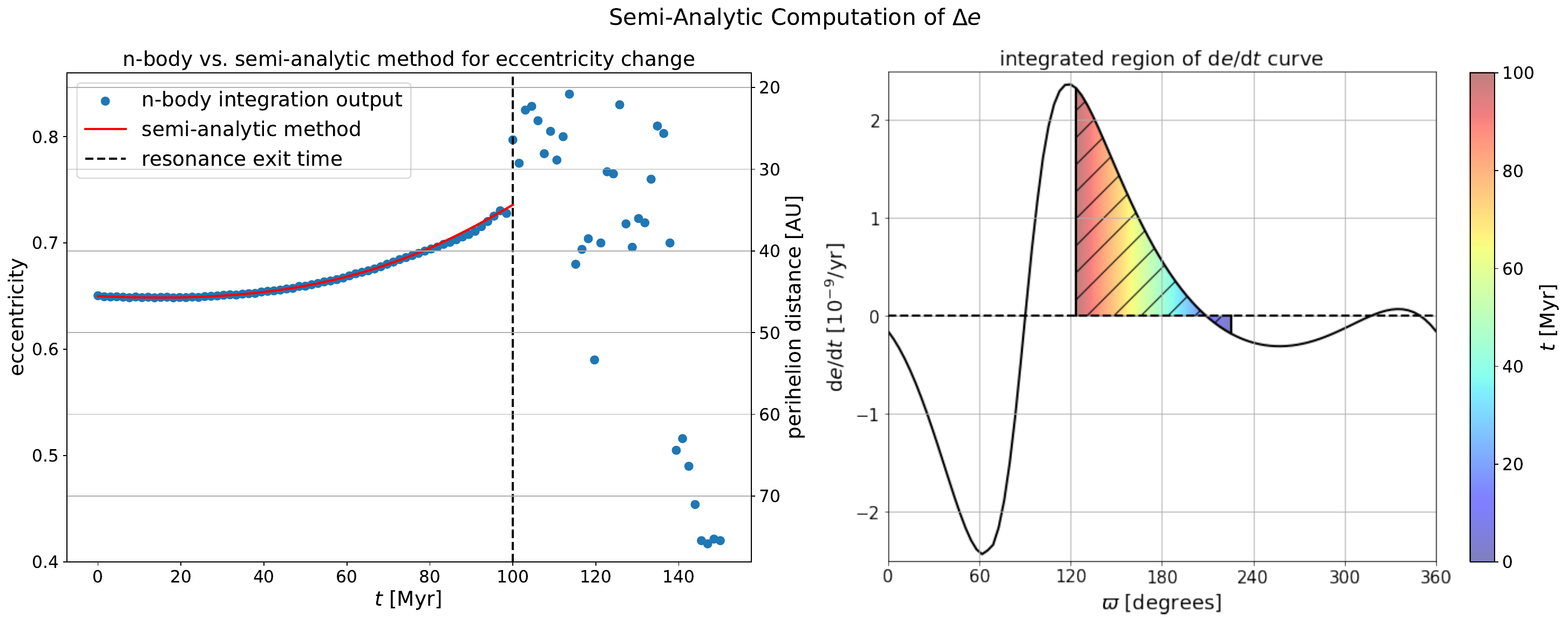}
    \caption{Illustration of the approximate, semi-analytic method for computing $\Delta e$ during a precession through one of the $\textup{d}e/\textup{d}t$ peaks. (Left): comparison of calculated eccentricity evolution (red curve) and $n$-body integration output (blue dots) over 150 Myr. Note that the perihelion distance ticks on the right side of this panel assume $a$ = 130.1 AU, which only holds while the object is still in resonance prior to $t=100$ Myr. (Right): Illustration of the integrated portion of the $\textup{d}e/\textup{d}t$ curve; The object starts at $\varpi = 225^{\circ}$, and exits resonance after 100 Myr at $\varpi = 123^{\circ}$. The integral (eq. 6) was performed assuming a constant $\varpi$ precession rate that was obtained from the $n$-body integration.}
    \label{fig:SemiAnalytic}
\end{figure}

\begin{figure}[ht]
    \centering
    \hspace*{-0.75cm}\includegraphics[scale = 0.45]{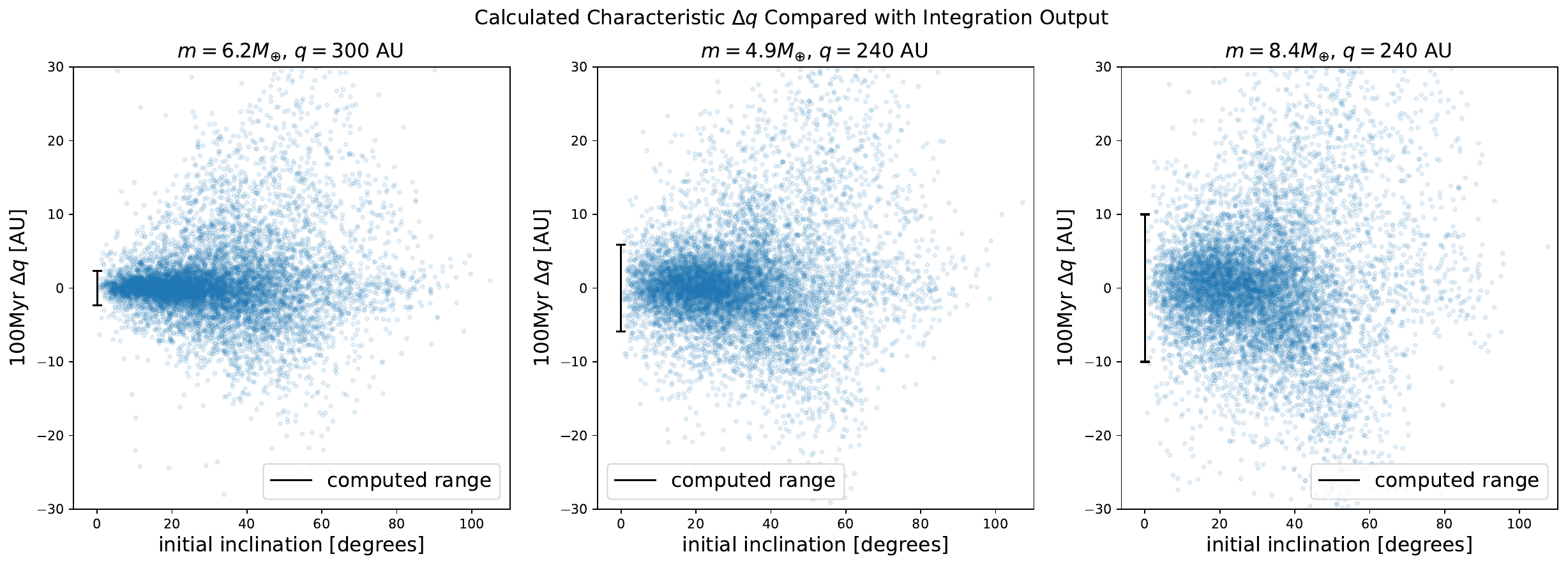}
    \caption{Calculated characteristic perihelion deviation range (black bar) versus actual perihelion change from $n$-body integration output (blue dots) for all integrated objects. (Far left): Control Simulation. (Center): $m = 6.2M_{\oplus}$, $q = 240$ AU perturber, aggregated over all 9 eccentricity-inclination combinations. (Far right): $m = 8.4M_{\oplus}$, $q = 240$ AU perturber, aggregated over all 9 eccentricity-inclination combinations. $\Delta q$ values were calculated using the $e = 0.3$, $i = 16^{\circ}$ planet for each $m$-$q$ combination shown, with the process outlined in Section~\ref{sec:arbitrary}.}
    \label{fig:deltaq_size}
\end{figure}

Consider a simple model containing only three bodies: The Sun, a zero-inclination test particle with $a = 130.1$ AU and $q = 45$ AU (our prototypical highly-stable 9:1 ``resonator"), and a Planet-9 like body with $m = 8.4M_{\oplus}$, $a = 400$ AU, $q = 240$ AU, $i = 16^{\circ}$, $\omega = 150^{\circ}$, $\Omega = 98^{\circ}$, which was one of the planets we used in our integrations. Over secular timescales, the test particle will experience a net torque from the planet that will slowly change its eccentricity while leaving its semi-major axis constant. This effect can be calculated using equation (5), either using numerical methods or by the series expansion approach outlined by \citet{MD1999}. The magnitude and sign of this interaction will depend on the orientation of the test particle's orbit (that is, on its longitude of perihelion, $\varpi$). The left panel of Figure~\ref{fig:dedt_curve} shows the result of this calculation as a function of $\varpi$; Note the large peaks where $\textup{d}e/\textup{d}t$ reaches values of order $10^{-9}$ / yr. The height of these peaks emerges as a quantity of interest, since test particle with a $\varpi$ value corresponding to the large positive peak would experience a perihelion change of $\Delta q = -10$ AU in only a few tens of millions of years, more than large enough to drive it into the scattering region.
From this simple model, one might expect the presence of an additional planet to cut large gaps into the $\varpi$ distribution of the resonant population, at locations that correspond to peaks in the curve. This simplified model neglects key dynamical effects, however, and to recover them we must add Neptune and the other giant planets into the picture. In doing this, one finds that the test particle now experiences a long-term \emph{precession} of $\varpi$ due to its resonant coupling with Neptune. The magnitude and sign of the induced eccentricity and inclination time derivatives now rapidly change as the orbit precesses, leading to chaotic oscillations of $e$ and $i$. To assign a characteristic size to the $e$ oscillations, we proceed by approximating $e$ and $i$ as constant over small intervals of precession, treating the particle as instantaneously ``sampling" the different values of $\textup{d}e/\textup{d}t$ from Figure~\ref{fig:dedt_curve} (left) as its longitude of perihelion changes. Now, the quantity of interest becomes total \emph{integrated} change in $e$ as the particle's longitude of perihelion precesses through the peaks of the $\textup{d}e/\textup{d}t$ curve. If the change is large enough to drive our would-be stable object into the scattering region, then the existence of a large, long-term stable 9:1 population may cast doubt on the existence of the planet from which this $\textup{d}e/\textup{d}t$ curve was computed. Note that in order to compute this integral, we must know the resonator's $\varpi$ precession rate, as the change in eccentricity is given by 
\begin{equation}
    \Delta e_{\textup{peak}} = \int_{\textup{peak}} \frac{\textup{d}e}{\textup{d}t}\frac{\textup{d}t}{\textup{d}\varpi}     \textup{d} \varpi \,.
    \label{eqn:delta-e}
\end{equation}
Our method is simply to assume a constant precession rate (which we obtain from a short $N$-body integration) and evaluate Equation~(\ref{eqn:delta-e}) using numerical integration. We neglect any effect of mean-motion resonances with the perturber, taking the secular average of the disturbing function over both mean longitudes in Equation \ref{eqn:dedt}. Figure~\ref{fig:SemiAnalytic} illustrates this method in action, showing the computation of $\Delta e$ over a single precession through the large peak in Figure~\ref{fig:dedt_curve} (left), and validates this approximate approach by comparing the result to the output of an $n$-body integration that includes the perturber, a test particle with initial values of $q=45$ AU and $\varpi=225^{\circ}$, and all four giant planets.
We note that this method yields mixed results in its effectiveness at accurately computing the eccentricity evolution of individual objects, especially over timescales longer than a few tens of millions of years. This uncertainty arises because the assumption of constant inclination and $\varpi$ precession rate does not hold in general. If the resonator's inclination and precession rates remain roughly constant, as in Figure~\ref{fig:SemiAnalytic}, then the computed trajectory will match closely with the output of an $N$-body integration, but in many cases this is not so. Nonetheless, the utility of this method in accurately predicting $N$-body dynamics is irrelevant, for we do not care if this approach over- or under-estimates the true eccentricity change in \emph{particular} cases; we aim only to assign a characteristic size to eccentricity oscillations that, over long timescales, are chaotic by nature. The effectiveness of our method toward this goal is well-illustrated in Figure~\ref{fig:deltaq_size}, which compares the computed characteristic $\Delta q$ for several of the planets we tested to the output of our integrations.

This procedure can be easily be carried out for an arbitrary planet and resonance, so long as one is able to obtain knowledge of the $\varpi$ precession rate. We should note, however, that while the choice of a zero-inclination resonator for computing the $\textup{d}e/\textup{d}t$ curves simplifies the method greatly, it is not always appropriate. If the planet in question has zero inclination, for instance, this method will tend to underestimate the expected action, as an alignment of the orbital planes eliminates one of the asymmetries that is responsible for inducing secular changes in eccentricity. Indeed, if this hypothetical planet were to also have zero eccentricity, this method would predict \emph{zero} action. In such cases, it is insightful to examine planes other than the ecliptic using a similar process, computing $\textup{d}e/\textup{d}t$ as a function of $\varpi$ for a particular value of $i$ and $\Omega$. This is done in the left panel of Figure~\ref{fig:dedt_curve} for a 9:1 resonator with $i = 5^{\circ}$ in the presence of a circular, zero-inclination planet with $m = 8.4M_{\oplus}$ and $a$ = 240 AU (the symmetry of this case makes the resonator's $\Omega$ unimportant). Even for a small inclination of $5^{\circ}$, the peaks of the curve reach similar heights to those in the right panel; clearly a large amount of action is still to be expected from this planet. When considering Planet 9-like perturbers with $i \sim 16^{\circ}$, though, the choice of zero inclination \emph{is} appropriate, as objects in the ecliptic will then exist at reasonable relative inclinations to the planet's orbital plane.

With this, we have a rudimentary framework for checking whether a given planet poses a risk to the existence of Gyr-stable objects in a given $n$:1 resonance with Neptune: If the relevant $\textup{d}e/\textup{d}t$ curve contains peaks that, when integrated assuming typical precession rates, induce changes in $q$ large enough to drive objects into the scattering region, then the existence of such objects casts doubt on that particular planet.

\begin{figure}[ht]
    \centering
    \hspace*{-0.75cm}\includegraphics[scale = 0.385]{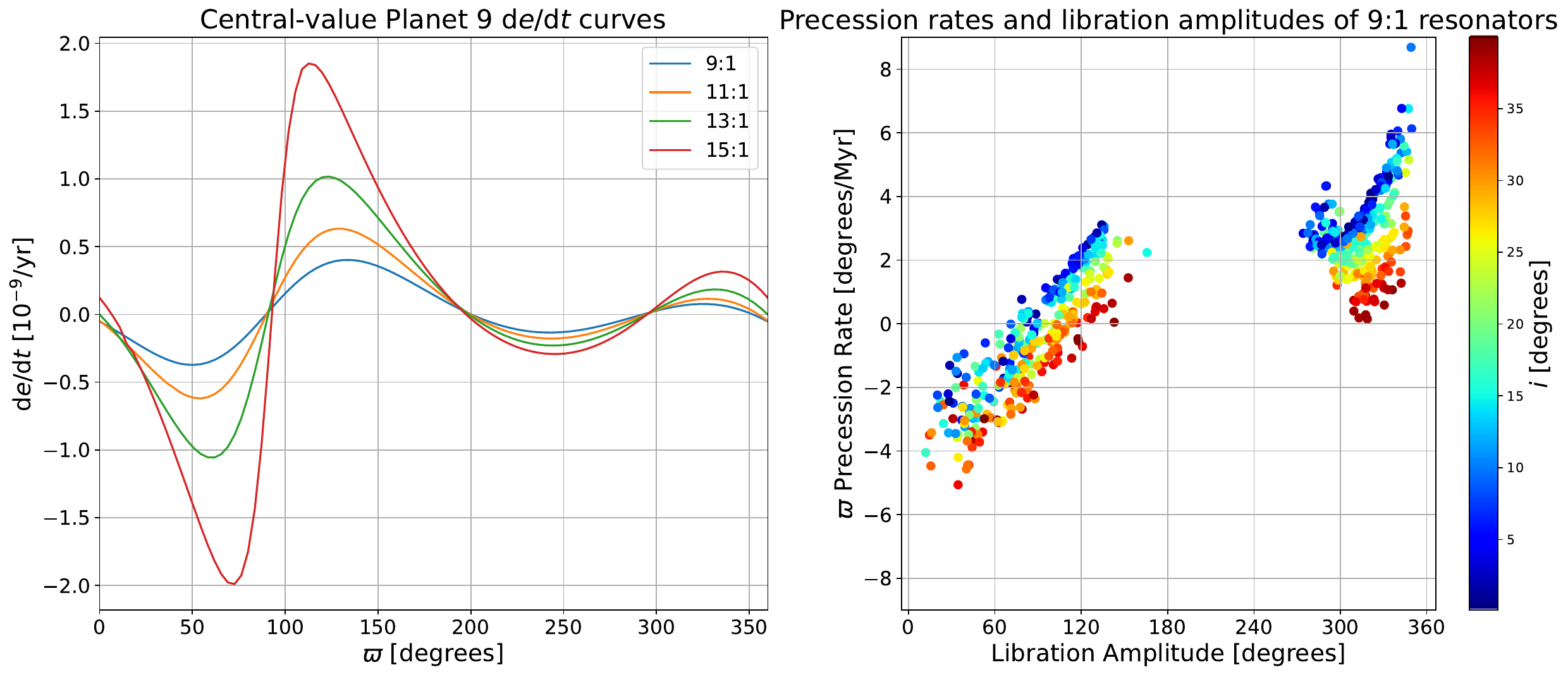}
    \caption{(Left): Computed $\textup{d}e/\textup{d}t$ curves for the central-value Planet 9 claimed by \citet{BB2022}, for four different resonances. For each resonance, a zero-inclination object with $q = 45$ AU was used. (Right): Libration amplitudes and precession rates of the generated 9:1 resonators that we used to determine the appropriate mean precession rate to use when calculating $\Delta q$ over the peaks in the curves shown in the left panel.}
    \label{fig:CVP9}
\end{figure}

As a final example, we apply this method to the central-value Planet 9 ($m = 6.2M_{\oplus}$, $a = 460$ AU, $q = 340$ AU, $i = 16^{\circ}$, $\omega = 150^{\circ}$, $\Omega = 97^{\circ}$) given by \citet{BB2022}, checking the danger posed to objects in several different resonances. For each resonance, we used an object with $q = 45$ AU and a semi-major axis exactly at the resonance to compute the $\textup{d}e/\textup{d}t$ curves, and integrated over the largest peak to compute $\Delta q$. To determine the precession rate to use for each resonance, we generated 1000 objects in each resonance in a similar manner to the process used for the 9:1 resonance in Section~\ref{sec:methods}, and took the mean of the absolute value of the precession rates of all objects. The resulting $\textup{d}e/\textup{d}t$ curves are shown in Figure~\ref{fig:CVP9}, and the computed $\Delta q$ values are shown in Table~\ref{tab:resonances}. Note in particular the value of $\Delta q$ = 14.7 AU for the 15:1. This increment is higher than the $\Delta q$ = 10 AU value for 9:1 resonators that we calculated using the $m = 8.4M_{\oplus}$, $q = 240$ AU planet. In other words, objects in the 15:1 resonance are likely to be disrupted. It follows that the discovery of a large, primordial population in the 15:1 resonance could potentially be used to constrain the existence of the aforementioned central-value Planet 9.
\begin{table}[ht]
    \centering
    \begin{tabular}{|c|c|c|c|}
    \hline
    \textbf{Resonance} & \textbf{Semi-Major Axis} [AU] & \textbf{Precession Rate} [degrees/Myr] & $\mathbf{\Delta} \mathbf{q}$ [AU]
    \\
    \hline
    \textbf{9:1} & 130.1 & 2.45 & 1.12 
    \\
    \hline
    \textbf{11:1} & 148.7 & 1.81 & 2.80
    \\
    \hline
    \textbf{13:1} & 166.3 & 1.41 & 6.51
    \\
    \hline
    \textbf{15:1} & 182.9 & 1.17 & 14.7
    \\
    \hline
    \end{tabular}
    \caption{Semi-major axis, $\varpi$ precession rate, and peak-integrated change in perihelion distance for multiple Neptune $n$:1 resonances under the influence of the central-value Planet 9 from (Brown \& Batygin 2022)}
    \label{tab:resonances}
\end{table}

As an aside, it is interesting to examine what factors determine the precession rate, as so far we have neglected to discuss this. Table~\ref{tab:resonances} shows that the rate tends to decrease for more distant $n$:1 resonances. Within a particular resonance, however, the precession rate depends on a large number of factors, including perihelion distance, inclination, and libration amplitude. Of particular note is the fact that the rate is typically much slower for objects in the inner libration islands, a feature illustrated in the right panel of Figure~\ref{fig:CVP9}. We hypothesize that this effect is responsible for the preferential erosion of the inner islands shown in Figure~\ref{fig:heatmaps}. The slow precession rate means that these objects linger in the $\textup{d}e/\textup{d}t$ peaks for longer times, and experience greater increases in eccentricity. 

\section{Summary and Conclusion}
\label{sec:conlude} 
This paper uses computational methods to study the Gyr-timescale stability and dynamics of synthetic objects in the 9:1 mean-motion resonance with Neptune in the presence of distant super-Earths. For the canonical Solar System, we found that the stability of resonators is highly sensitive to perihelion distance (Figure~\ref{fig:survival}). Objects with $q \lesssim 40$ AU are rapidly ejected from resonance via planetary scattering with Neptune, while objects with $q \gtrsim 40$~AU are much more likely to remain in resonance for Gyr timescales. We observed long-term stability across a broad range of inclinations, although a mild instability arises for objects with inclinations between $30^{\circ}$ and 60$^\circ$.

Planet 9-like perturbers caused a significant decrease in the fraction of resonators that survive the 1~Gyr simulation within 1~AU of the 9:1 resonance. The most aggressive perturbers under consideration ($m = 8.4M_{\oplus}$, $q = 240$ AU) reduced the number of surviving objects by slightly more than half when compared to the canonical Solar System, while all other planets induced only a small decrease (Figure~\ref{fig:classification}). We also observed a preferential ejection of low-amplitude resonators, a phenomenon which we attribute to their slower $\varpi$ precession rates (Figures~\ref{fig:heatmaps} and  \ref{fig:CVP9}). Our simulations show that these planets destabilize resonators by inducing secular changes in eccentricity, which can drive the perihelion distances to low values where they are ejected by Neptune via scattering (Figure~\ref{fig:q_kickout}). These results can be extrapolated to include general planets and resonances, and we have constructed a heuristic method for quantifying the characteristic size of the induced eccentricity changes for a given planet and resonant population (Section~\ref{sec:arbitrary}). Applying this method to the central-value Planet 9 from \citet{BB2022}, we find that this planet would cause long-term instability in resonant populations at or beyond the 15:1 resonance (semi-major axis $a \approx 183$ AU).

Our findings make clear that the two 9:1 resonators discussed in \citet{Volk2018} cannot be used to place meaningful constraints on the Planet 9 hypothesis, as the only planets which caused a significant depletion over 1 Gyr (those with $m = 8.4M_{\oplus}$ and $q = 240$ AU) are considered unrealistic in the current literature \citep{BB2021, BB2022}. Additionally, it is not yet clear whether the objects discovered by OSSOS are of a primordial origin, or if they became transiently stuck in resonance during the last $\sim$1 Gyr. However, the upcoming completion of next-generation telescopes will soon enable the discovery of many more objects in Neptune's distant $n$:1 resonances. The Vera Rubin Observatory is expected to increase the number of known TNOs by 1-2 orders of magnitude through its ten-year Legacy Survey of Space and Time \citep{Jones2016}, though the discovery of a large number of highly distant objects may require a deeper target survey as discussed by \citet{Lawler2019}. This present work provides a foundation for the analysis and interpretation of upcoming discoveries, including constraints on possible additional planets in the Solar System. 

\appendix

\section{The \texttt{WHFast} Integrator}
We make use of \texttt{REBOUND}'s \texttt{WHFAST} integrator, described in \citep{WHFast}, for all of the integrations analyzed in this work. While this symplectic Wisdom-Holman integrator is highly CPU-efficient, it may produce to inaccuracies in situations involving close encounters with massive bodies. To accurately model such scenarios, it is instead necessary to employ an adaptive integrator such as \texttt{REBOUND}'s \texttt{IAS15}, described in \citep{ias15}. Use of WHFast is a natural choice for studies of resonant TNO dynamics, as the geometry of resonance generally acts to protect test particles from close encounters with Neptune. This integrator may not be appropriate for the discussion in Section~\ref{sec:adiff}, however, as particles which have left the 9:1 resonance are no longer protected from close encounters. Since many of these particles have perihelia approaching Neptune's semi-major axis ($a=30$ AU), the use of WHFast requires additional justification in this context.

To alleviate this concern, we re-ran the integrations described in Section~\ref{sec:methods} for the perturber with $m=8.4M_\oplus$, $q = 240$ AU, $e = 0.3$, and $i = 16^{\circ}$ using the hybrid \texttt{MERCURIUS} integrator \citep{Mercurius}, which switches to \texttt{IAS15}'s high-order, adaptive step-size routine when a close encounter occurs. As expected, the fraction of objects which survive for 1 Gyr as librating resonators is not statistically different in the \texttt{MERCURIUS} integration (0.083 +/- 0.0091) when compared to the previous \texttt{WHFAST} result (0.093 +/- 0.0096). Additionally, no meaningful difference was found in the final distribution of any orbital element, or in the fraction of objects ejected from the Solar System (0.261 +/- 0.016 for \texttt{MERCURIUS} vs. 0.246 +/- 0.016 for \texttt{WHFAST}). For the analysis discussed in Section~\ref{sec:adiff} and illustrated in Figures~\ref{fig:classification} and~\ref{fig:a_vs_e}, the \texttt{MERCURIUS} and \texttt{WHFAST} results can thus be considered equivalent.

\section{Mean-Motion Resonances with the Perturber}
Much of the analysis in this work is carried out by aggregating the results of several similar distant perturbers into a single unit. The validity of this approach may be compromised by the fact that several of the perturbers tested are near simple-integer period ratios with the test particles in Neptune's 9:1 MMR, while others are not. In particular, the semi-major axis corresponding to Neptune's 9:1 MMR, $a = 130.1$ AU, is roughly 0.1 AU distant from the 2:7 MMR with the ($q = 240$ AU, $e = 0.2$) perturbers, 1 AU distant from the 1:6 MMR with the ($q = 300$ AU, $e = 0.3$) perturbers, and 2 AU distant from the 1:11 MMR with the ($q = 385$ AU, $e = 0.4$) perturbers. To demonstrate the validity of this analysis, it is important to show that such proximity to the perturber's MMRs does not appreciably alter the strength or character of the dynamics at play.

It is straightforward to show that in the cases tested, the perturber's sunward resonances do not overpower the the 9:1 resonant libration with Neptune. To accomplish this, we ran additional 10 Myr extension integrations on all non-ejected particles for each of the near-resonant perturbers with $m = 8.4M_{\oplus}$ and $i = 16^{\circ}$. Each test particle was then checked for libration in both of the relevant resonant angles: $\phi_N = 9\lambda - \lambda_N - 8\varpi$ for the 9:1 resonance with Neptune, and $\phi_P = n\lambda_P - m\lambda - (n-m)\varpi_P$ for the $m$:$n$ resonance with the perturber. Here the subscript $N$ refers to Neptune, while the subscript $P$ refers to the perturber. The resulting resonant demographics are summarized in Table \ref{tab:resonances}.

\begin{table}[ht]
    \centering
    \begin{tabular}{|c|c|c|c|}
    \hline
    \textbf{Perturber} & \textbf{Perturber} & \textbf{\# Librating in $\phi_N$} & \textbf{\# Librating in $\phi_P$}
    \\
      & \textbf{Resonance} & \textbf{9:1 Resonators} & 
    \\
    \hline
    $q = 240$ AU, $e = 0.2$ & 2:7 & 63 & 4
    \\
    \hline
    $q = 300$ AU, $e = 0.3$ & 1:6 & 242 & 7
    \\
    \hline
    $q = 385$ AU, $e = 0.4$ & 1:11 & 401 & 2
    \\
    \hline
    \end{tabular}
    \caption{Number of candidate resonators, $\phi_N$ librators, and $\phi_P$ librators at the end of the 2 Myr extension integrations with near-resonant perturbers. For each ($q, e$) combination, the extension integrations were performed on all non-ejected objects from the simulations involving the $8.4M_{\oplus}$, $i = 16^\circ$ perturber with the indicated ($q, e$) values. To be classified as a librator, the object satisfy max($\phi$) - min($\phi$) $\leq 355^{\circ}$.}
    \label{tab:perturber_res_demographics}
\end{table}

The $q = 240$ AU, $e=0.2$ perturbers are of particular interest here. This $q$-$e$ combination produces the smallest semi-major axis of all tested perturbers, and in this configuration the alignment between Neptune's 9:1 resonance and the perturber's 2:7 resonance is virtually exact. To verify that this resonance overlap does not effect our results, we ran an additional 1,000 objects through a 1 Gyr integration with an $8.4M_{\oplus}$, $q = 240$ AU, $e = 0.23$, $i = 16^\circ$ perturber, and compared the results to the $8.4M_{\oplus}$, $q = 240$ AU, $e = 0.2$, $i = 16^\circ$ perturber. This small shift in eccentricity corresponds to a change in semi-major axis of roughly 11 AU, eliminating the resonance overlap without drastically changing the perturber's orbit. In both cases, similar number of the 1,000 objects were found to exhibit libration in Neptune's 9:1 during the 2 Myr extension integrations (73 objects for $e = 0.23$ versus 63 objects for $e = 0.2$), and no significant differences were observed in the final orbital element distributions of non-ejected objects. Lastly, the distribution of final 9:1 libration amplitudes did not differ in either case. It is clear from this that proximity to an MMR with the perturber does not significantly affect the conclusions of our analysis.

\hfill

This material is based upon work supported by the National Aeronautics and Space Administration under grant No.\ NNX17AF21G issued through the SSO Planetary Astronomy Program and by the National Science Foundation under grant No.\ AST-2009096. This research was supported in part through computational resources and services provided by Advanced Research Computing at the University of Michigan, Ann Arbor. MWP would like to thank Christina Porter for their support throughout this work.

%\bibliography{sample631}{}
%\bibliographystyle{aasjournal}

\end{document}